\documentclass{emulateapj}

\newcommand{\be}{\begin{equation}}
\newcommand{\ee}{\end{equation}}
\newcommand{\beqn}{\begin{eqnarray}}
\newcommand{\eeqn}{\end{eqnarray}}
\newcommand{\bld}[1]{\mbox{\boldmath$#1$\unboldmath}}

\begin{document}

\title{Self-Similar Solutions of Triaxial Dark Matter Halos}
\author{Yoram Lithwick\altaffilmark{1,2} and Neal Dalal\altaffilmark{2}}

\altaffiltext{1}{Department of Physics and Astronomy, Northwestern University, 2145 Sheridan Rd., Evanston, IL 60208}
\altaffiltext{2}{Canadian Institute for Theoretical Astrophysics, 60 St. George Street, Toronto, ON M5S 3H8, Canada}

\begin{abstract}
We investigate the collapse and internal structure of dark matter halos.
We consider halo formation from initially scale-free perturbations, 
for which gravitational collapse is self-similar. 
Fillmore and Goldreich (1984) and Bertschinger (1985)
solved the one dimensional (i.e. spherically symmetric) case. 
We generalize their results 
by formulating
the {\it  three dimensional} self-similar equations.
We
solve the equations numerically and analyze
the similarity solutions in detail, focusing on the internal 
density profiles of the collapsed halos.
By decomposing the total density into  
subprofiles of particles that collapse
coevally, we identify
two effects as the main determinants
of the internal density structure of halos:
 adiabatic contraction and 
 the shape of a subprofile shortly after collapse;
 the latter largely reflects the triaxiality of the subprofile.
We develop a simple model that describes
the results of our 3D simulations.
In a companion paper, we apply this model
to more realistic cosmological fluctuations, and
thereby
explain the origin of the nearly universal (NFW-like)
density profiles  found
in N-body simulations.

\end{abstract}

\section{Introduction}

As the Universe expands, small fluctuations in the dark matter density
grow in amplitude.  Eventually, these dark matter fluctuations grow
nonlinear, leading to collapse and virialization into structures
termed halos.  The internal structure of dark matter halos has long
been a topic of interest.  Halo structure significantly affects galaxy
formation and evolution, since galaxies form at the centers of
halos. It can also have implications for dark matter detection, which
depends strongly on the dark matter density inside collapsed objects.

N-body simulations have shown that halos forming in
hierarchical, Cold Dark Matter (CDM) cosmologies have fairly generic
density profiles. 
Navarro, Frenk, \& White (``NFW,'' 1996, 1997)
 suggested that CDM halos
have {\it universal} profiles, with $\rho\propto r^{-3}$ at large radii and
$\rho\propto r^{-1}$ at small radii.  Subsequent numerical work found
qualitatively similar results \citep[e.g.][]{Moore98,ViaLactea1,
Gao08,
GHalo,
ViaLactea2,
NLSetal10},
although the precise
shape of the innermost profile is still under debate.

It has proven a long-standing puzzle to explain why halos
are well described by the NFW profile.
Identifying the important processes using
conventional cosmological N-body simulations has been challenging, in
part because of the limited dynamic range of such simulations.  If the
mechanism setting halo profiles is truly generic, however, then we can
investigate it using particular test cases that can be studied in great
detail.  One standard technique for achieving high resolution is to
study problems that admit self-similar solutions, which 
is the approach we follow in this paper.

\cite{FG84} and \cite{Bert85}  calculated the self-similar
collapse of spherically symmetric profiles.
They considered the evolution of a  linear overdensity
of the form
$\delta_{\rm lin}\propto r^{-\gamma}$ 
in a flat CDM Universe.
This problem has only a single characteristic lengthscale, $r_*$, 
which is where the overdensity is
of order unity.  Inside of $r_*$ a virialized
halo forms, and outside of $r_*$ density perturbations remain small and
linear.  As time proceeds, the virialized halo grows by accreting
from its surroundings, and $r_*$ increases.
Yet the 
properties of the halo and its surroundings are independent
of time when scaled relative to $r_*$.
The interior density 
of the virialized halo scales as a power-law in radius, 
$\rho\propto r^{-g}$, where 
$g= 2$ for $\gamma\leq 2$, and
$g= 3\gamma/(1+\gamma)$
for $\gamma\geq 2$ \citep{FG84}.
 Hence the interior 
slope is always much steeper than the interior NFW slope.

Subsequent work expanded upon this analysis.  \citet{Ryden93}
calculated the self-similar collapse of axisymmetric
profiles.\footnote{\cite{DuffySikivie}
also show that self-similarity does not require spherical
symmetry.}
 She showed that $\gamma=2$ yielded an inner
slope of $-g\simeq -2$, in accord with the spherically symmetric prediction.
But $\gamma=1$ yielded $-g\approx -1.5$.
Surprisingly, the work of  \cite{Ryden93} has not been followed up
in the literature.
 In particular, it has not yet been extended to three dimensions,
which is the focus of the present paper.

As we describe below, a number of authors
have used qualitative arguments, as well 
as model equations, to argue 
that self-similar, triaxial collapse should produce asymptotic inner
slopes given by the Fillmore-Goldreich expression
$g=3\gamma/(1+\gamma)$ 
for all $\gamma$.
\citep[e.g.][]{WhiteZaritsky92,Nusser01,Ascasibar04,Ascasibar07}.
In this paper, we shall confirm with our 
solutions of the exact equations 
that this indeed holds true, although in some cases
this slope is only reached 
on scales  $\lesssim 10^{-5}$ of the virial
radius.

More recently,  \cite{VMW10} carried out N-body simulations
initialized with the spherically symmetric initial 
conditions of \cite{FG84}.  They found that even though the initial
conditions are spherically symmetric, the radial orbit instability
acts to make the evolution non-spherical.
 \cite{ZB10} considered spherical self-similar solutions, but
 with a prescription based on tidal torques for generating non-radial
 velocities.

In this paper, we consider the evolution of a linear overdensity
of the form $\delta_{\rm lin}=f(\theta,\phi)r^{-\gamma}$, where
$f$ is an arbitrary function of zenith and azimuth.
We emphasize that we solve the exact evolutionary equations, 
and our solutions are exact, subject only to numerical errors.
In this sense, our nonspherical solutions differ from previous
models of nonspherical collapse \citep[e.g.,][]{el,BondMyers96}
that resorted to model equations.

The paper is structured as follows.  In \S 2, we describe our method
for obtaining the self-similar solutions to the equations of motion
governing the collapse of initially scale-free perturbations.  In \S
3, we introduce an oversimplified model---the ``frozen model''---that
will prove helpful in interpreting the results of the full simulations.
In \S 4, we review the spherically symmetric solutions of 
\cite{FG84,Bert85}.  In \S 5, we present the solution to axisymmetric
collapse.  In \S 6, the heart of this paper, we present a suite
of 3D self-similar solutions, and analyze them in detail, 
with the primary goal of isolating the physics that is responsible 
for the interior density profiles. 
In \S 7, we introduce a toy model that incorporates the mechanisms
that we identified in \S 6.  
In  \cite{Paper2}
 we apply the
toy model to more realistic cosmological fluctuations, and show that
NFW-like profiles are a generic outcome of the dissipationless
collapse of peaks of Gaussian random fields.

\section{Self-Similar Equations and Method of Solution}

\subsection{Equations of Motion}

We consider a region of space in a flat CDM Universe that has linear overdensity
\be
\delta_{\rm lin}(\bld{r}) = {\rm const}\times r^{-\gamma}f(\theta,\phi) 
\label{eq:dlin}
\ee
at a fixed time, 
where $\bld{r}$ is the proper
radial displacement from the origin,  $\gamma$ is a constant and $f$ is an arbitrary function of zenith and azimuth.
As long as the overdensity satisfies $|\delta|\ll 1$, linear perturbation theory implies that
\beqn
\delta(t, \bld{r}) &=&  t^{2/3}\left( {r\over t^{2/3}} \right)^{-\gamma}f(\theta,\phi)  
\label{eq:lin1}
\\
&=& \left( {r\over r_*(t)} \right)^{-\gamma}f(\theta,\phi) \ , \ \ \ ({\rm for}\ \delta\ll 1)
\label{eq:lin2}
\eeqn
where 
\be
r_*(t)\equiv t^{{2\over 3}+{2\over 3\gamma}} \ ,
\ee
and we have absorbed a constant prefactor into $f$.
The quantity $r_*(t)$ is, aside from a multiplicative constant, equal to the
 characteristic lengthscale at which nonlinearities become important ($\delta\sim 1$).  
As we now show, the evolution is self-similar when all lengthscales are
scaled to $r_*$.

A particle's equations of motion are
\beqn
{d{\bld r}\over dt} &=& {\bld v} \\
{d{\bld v}\over dt}&=& -\bld{\nabla_r}\phi \ ,
\label{eq:r2dot}
\eeqn
where ${\bld r}$ and ${\bld v}$ are its position and velocity, 
and $\phi$ is the gravitational potential (not to be confused with 
the azimuthal
angle).
Different particles may be labelled by their positions $\bld{r}_i$ at an initial time
$t_i$, 
\beqn
\bld{r} = \bld{r}(t, {\bld{r}_i}) \ , \ \bld{v} = \bld{v}(t, {\bld{r}_i}) \ ,
\label{eq:ii}
\eeqn
where $t_i$ is chosen to be sufficiently small that all particles 
of interest expand with the Hubble flow at that time.

Scaling  lengths with $r_*$ and times with $t$, we
define the scaled radius and velocity $(\bld{R},\bld{V})$ via\footnote{
Our convention is to denote scaled variables by capitalized letters, 
with scaled variables equal to unscaled ones at $t=1$; the 
independent variable
$s$ is the exception to this rule.
}
\beqn
\bld{r}(t,\bld{r}_i) &=& r_*\bld{R}(s,\theta_i,\phi_i) \label{eq:rscal} \\
\bld{v}(t,\bld{r}_i) &=& {r_*\over t}\bld{V}(s,\theta_i,\phi_i)  \ ,
\eeqn
where
\be
s\equiv{r_i\over r_*}\left(
{t\over t_i}
\right)^{2/3} \ 
\ee
is the scaled initial comoving radius.
We shall employ $s$
as our independent radial variable; it
 is a Lagrangian co-ordinate that labels particles by their
initial radius.
Its form may be understood as follows.
At early times a particle expands with the Hubble flow, and hence its
$r_i$ depends on $t_i$ via
  $r_i\propto t_i^{2/3}$. Therefore  different particles may be labelled by their value of $r_i/t_i^{2/3}$.  Scaling $r_i$ with 
$r_*$ and $t_i$ with $t$ leads to our form for $s$. 
Equivalently, each particle may be labelled by the mass it initially enclosed
as it expanded
 with the Hubble flow, $m_i=(4\pi/3)\bar{\rho}(t_i)r_i^3$, 
where $\bar{\rho}=1/(6\pi t^2)$  is the background density of
a homogeneous flat Universe.\footnote{
We set the gravitational constant $G$ to unity throughout this paper.  
One may restore it by multiplying  all quantities with dimensions of mass  
by $G$.
}
Therefore its scaled initial mass
\be
M_i\equiv {m_i\over r_*^3/t^2}={2\over 9}s^3 
\label{eq:mi}
\ee
is a simple function of $s$.

Since the gravitational potential has units of velocity squared, we define the scaled
potential $\Phi$ via
\be
\phi(t,{\bld r}) \equiv {r_*^2\over t^2}\Phi(\bld{R}) \ .
\ee
\newline
The  equations of motion then become
\be
{d\over d\ln s}
\left(\begin{array}{c}{\bld R} \\ {\bld V}\end{array}\right)=
\left(\begin{array}{cc}
\gamma+1
& -{3\gamma\over 2} \\
0 & 1-{\gamma\over 2}\end{array}\right)
\left(\begin{array}{c}{\bld R} \\ {\bld V}\end{array}\right) 
+
{3\gamma\over 2}
\left(\begin{array}{c}0 \\ {\bld \nabla_R\Phi}\end{array}\right)
\label{eq:rv}
\ee

It remains to calculate $\Phi({\bld R})$.  Given the trajectories
 $\bld{r}(t,\bld{r}_i)$ for all particles, one can calculate the corresponding
mass density, and then invert Poisson's equation to yield 
the potential.
Consider the particles that at time $t_i$ lie near
 ${\bld r}_i$, within the volume element $d^3{\bld r_i}$.
The mass in these particles is
\beqn
dm&=&\rho(t_i,{\bld r}_i) d^3{\bld r_i} \\
&=& {1\over 6\pi  t_i^2}d^3{\bld r_i}
\eeqn
where $\rho$ is the mass density, which at time $t_i$ is given by 
the background density.
By mass conservation,  the density $\rho$ at position $\bld{r}=\bld{r}(t,\bld{r}_i)$
due to these particles is given by
\be
\rho(t,\bld{r} \vert \bld{r}_i)d^3\bld{r} = {1\over 6\pi  t_i^2}d^3{\bld r_i} \ . \label{eq:jac}
\ee
The third argument of $\rho$ (i.e. $\bld{r}_i$) is necessary because
particles with different vaues of $\bld{r}_i$ can contribute to the density
at position $\bld{r}$ due to shell crossing.  The total density 
is the sum of the above $\rho$ over all $\bld{r}_i$ such that
 $\bld{r}=\bld{r}(t,\bld{r}_i)$.
 
We denote the scaled density with a capital $\rho$ (=$P$), defined
via
\be
\rho(t,\bld{r} \vert \bld{r}_i) = {1\over t^2}P(\bld{R} \vert \bld{s}) \ , 
\ee
in which case the scaled version of equation (\ref{eq:jac})
is
\be
P(\bld{R}\vert \bld{s})d^3\bld{R} ={d^3{\bld s}\over 6\pi} \ ,  \label{eq:p}
\ee
where we have introduced the vector
\be
{\bld s} \equiv {{\bld r_i}\over r_*}\left(
{t\over t_i}
\right)^{2/3} \ .
\label{eq:svec}
\ee
The total density is the sum over all streams that contribute to $\bld{R}$:
\be
P({\bld R})=\sum_{\bld{s}: \bld{R}(\bld{s})=\bld{R} } P(\bld{R}\vert \bld{s}) \ .
\ee
Henceforth the symbol $P$ without an argument will denote the total 
density $P(\bld{R})$.

Given $P$, one can solve for $\Phi$ by inverting the scaled
Poisson equation
\be
\nabla_R^2\Phi = 4\pi P \ . \label{eq:poiss}
\ee

Equations (\ref{eq:rv}), (\ref{eq:p}), and (\ref{eq:poiss}) are the self-similar
equations. They form a closed system.
Note that  $t$, $r_*$, $t_i$, and $r_i$ only enter these equations in the 
 combination $s$.  
 The self-similar equations are exact, following directly
 from the definitions of $r_*$ and the scaled variables.  
 Therefore the evolution is self-similar when the linear density field
 can be written as a function of the scaled variables, i.e. when it has
 the form given by equation (\ref{eq:dlin}).

\subsection{Method of Solution}
\label{sec:mos}

We solve the self-similar equations iteratively.  
The scaled density
is first set to its linear value everywhere, i.e., to
\be
 P({\bld R})= P_{\rm lin}({\bld R})\equiv {1\over 6\pi } \left( 1+R^{-\gamma}f(\theta,\phi) \right)  \ ,
 \label{eq:linP}
 \ee
 where $f$ is any prescribed function and $0< \gamma < 3$.
Inverting Poisson's equation determines
 $\Phi({\bld R})$.  We use this $\Phi$ to integrate
the trajectories of particles.
Each particle
is initialized at large $s$, where it expands with the Hubble flow:
${\bld R_0} = {\bld s}, 
 {\bld V_0} = {2\over 3}{\bld s}$.  Its subsequent
 trajectory is determined by integrating equation (\ref{eq:rv})
 from large to small $s$ with fixed $\Phi(\bld{R})$.
 At each step $ds$ of this integration, we record how
 much the particle contributes
 to the local density  (eq. [\ref{eq:p}]).   Repeating this
 integration for particles at all initial $\theta$ and $\phi$
 produces the full
 density field $P({\bld R})$.    Inverting Poisson's equation
 then
  determines the new $\Phi({\bld R})$, 
 which is used in the second iteration step.
This cycle continues until successive $\Phi$'s agree with 
each other, which typically takes 10-20 iteration steps, depending
on the simulation parameters.
See \S \ref{sec:ssc} for a more detailed description of
our method.

\section{Frozen Model}
\label{sec:frozen}

At large radii, $r\gtrsim r_*(t)\equiv t^{{2\over 3}+{2\over 3\gamma}}$, the evolution is linear and  particles
nearly expand with the Hubble flow.    Eventually $r_*(t)$ will overtake
such particles, whereupon their  evolution becomes nonlinear 
and they gravitationally collapse.  In the present section, we ask what
the density profile would be if, instead of collapsing,
each particle's proper radius remained frozen as soon as it crossed
$r_*$.
This is essentially identical to the ``circular orbit'' model of
\citet{RydenGunn87}, in which mass shells are placed on circular
orbits with energy equalling that at turnaround.  Their circular-orbit
model freezes shells at 
half the turnaround radius, 
while our frozen model freezes
shells at $r_*$, and so the resulting profiles are identical up to an
overall rescaling.  The key feature is that both of these models avoid
shell crossing.  Although this frozen model is unphysical, it is
useful as a baseline to compare against the full evolution.

\begin{figure}
\centerline{\includegraphics[width=0.5\textwidth]{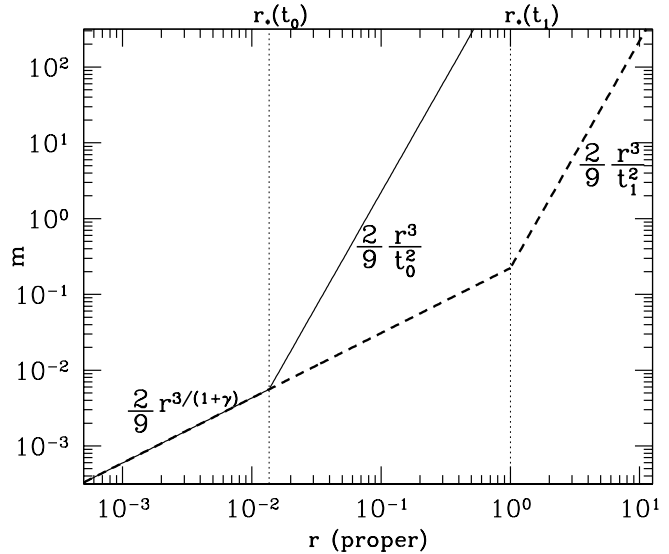}}
\caption{Frozen Model Mass Profile \label{fig:frozen}
($\gamma=2.5$): 
The solid line shows the enclosed mass profile $m(t,r)$ 
at time $t_0=0.01$, and the dashed line shows it at
time $t_1=1$.
 At large radii, $r>r_*(t)$, particles
expand with the Hubble flow, and the enclosed mass
is given by the background expression $(2/9)r^3/t^2$.
At small radii, particles are frozen.  Hence
the form of $r_*(t)$ dictates the profile at $r<r_*$.\newline
}
\end{figure}

 Figure \ref{fig:frozen} shows two snapshots
 of the enclosed mass profile,
 \be
m(t,r)\equiv 4\pi\int_0^r{\rho}(t,r)r^2dr \ , 
\ee
in this model.
At radii larger
than $r_*$, the enclosed mass profile is the product
of
the background density $=1/(6\pi t^2)$ and the volume of a sphere
of radius $r$, i.e., $m=(2/9)r^3/t^2$.  
At radii smaller than $r_*$,
the mass
profile is time-independent because of  the frozen assumption, and
thus 
$m=(2/9){r^3/ t_*(r)^2}$,
 where $t_*(r)$ is the inverse
of $r_*(t)$; i.e., the frozen mass profile is
\be
m=
   \cases{
      {2\over 9}r^{3/(1+\gamma)}, & if $r<r_*$\cr
      {2\over 9}{r^3\over t^2}, & if $r>r_*$\cr
      \label{eq:mf}
   }
\ee
It follows that the interior density profile
scales as 
\be
\rho\propto r^{-g_f} \ , 
\ee
where
\be
g_f\equiv {3\gamma\over\gamma+1} \ ;
\ee
we call $-g_f$ the frozen slope.

Real collapsing  solutions---in 1D as well as in 3D---are 
obviously more complicated than the frozen model.  
We distinguish two effects, either  of which may 
cause the interior spherically-averaged
 density profile to deviate from the frozen slope:
\begin{enumerate}
\item {Shell Profiles:}
Consider the set of particles that initially lie in a thin
spherical shell that expands with the Hubble flow.
In the frozen model, it is assumed that once
the radial co-ordinates of these shell particles\footnote{
Throughout this paper, we refer to particles that initially lie
within the same thin spherical shell as ``shell particles,''
because their Lagrangian co-ordinates occupy a shell.
}
cross $r_*$, they freeze;
hence they lay down a density profile that
vanishes everywhere except at the distance at
which they freeze.
By contrast, in the actual 
self-similar solutions, 
shell particles
lay down an extended density profile as they execute their orbits,
which we call the ``shell profile.''
The shape of the tail of the shell profile (i.e. for $r\rightarrow 0$)
can affect the total density profile at small radii.
\item {Shrinking Apoapses:}
The apoapses of shell particles can gradually
shrink in time, rather than the particles remaining frozen.
\end{enumerate}

 As we shall show, in general both shell profiles and shrinking apoapses
 are present in the full solutions.  But even when they do occur, they 
 do not necessarily cause the interior density profile to deviate from 
 the frozen slope at very small radii.  In particular, the total density slope will be $-g_f$
 as long as 
 \begin{enumerate}
 \item
  each shell tail remains shallower than 
 $-g_f$; and, 
 \item
 the apoapses eventually stop shrinking after 
a finite time
 \end{enumerate}
  
 \subsection{Lagrangian Variables $r_f$ and $t_f$}
 \label{sec:rftf}
 
When considering the evolution of particles in the self-similar solutions, 
it will prove convenient
to employ new Lagrangian variables based on the frozen model.
In place of
$t_i$ and $r_i$ (eq. [\ref{eq:ii}]), we define
   $r_f$
as the radius 
  at which a particle
with given initial $r_i$ and $t_i$ would freeze in the frozen model; similarly, 
we define $t_f$ as its time of freeze-out in the frozen model.
Since
$r_f=r_*(t_f)$ and
 $r_f/r_i=(t_f/t_i)^{2/3}$, we have
 \beqn
 {r_f}&\equiv&\left(r_i^3/t_i^{2}  \right)^{(1+\gamma)/3} \\
 t_f&\equiv&\left(r_i^3/t_i^{2} \right)^{\gamma/2} \ .
 \eeqn
Scaling relative to $r_*$ and $t$ yields
\be
{r_f\over r_*}=  s^{1+\gamma} \ , \ \ \ {t_f\over t}= s^{3\gamma/2} \ .
 \label{eq:rrf}
\ee

\section{Spherically Symmetric Solutions}
\label{sec:sss}

We review here the  well-known spherically symmetric
self-similar solutions \citep{FG84,Bert85}, 
focussing in particular on the interior density profile.
The linear density field is prescribed to be $\propto R^{-\gamma}$, 
independent of $\theta$ and $\phi$.
Note that 
our parameter $\gamma$ is related to
Fillmore \& Goldreich's  $\epsilon$ 
via $\gamma=3\epsilon$ for their spherically symmetric
solutions.  In these  solutions, all particles
have purely radial trajectories, and hence are forced to 
cross through the origin repeatedly after collapse.\footnote{
Numerous authors have considered  
spherically symmetric solutions in which particles are
assigned non-radial velocities with various prescriptions
 \citep[e.g.,][]{Nusser01,ZB10}.
However, since we solve the exact equations of motion, we must
consider departures from spherical symmetry
in order to generate non-radial motions.
}
We describe the spherical solution in great detail, in order to lay the
groundwork for our discussion of the more complicated triaxial
similarity solutions.

In spherically symmetric collapse, the density profile
for $R\ll 1$ scales as
\be
P\propto R^{-g} \ ,
\ee
where
\be
g =
   \cases{
      g_f\equiv 3\gamma/(1+\gamma), & for $\gamma>2$\cr
     2, & for $\gamma<2$\cr 
   }
\ee
\citep{FG84,Bert85},
i.e., the slope differs from the frozen slope
when $\gamma<2$.
To illustrate the reason for this  behavior, and to set the stage for the more complicated
non-spherical cases, we present two spherical simulations, with
$f=1$ in equation (\ref{eq:linP}):
one simulation with $\gamma>2$ and the other with $\gamma<2$.

\subsection{
Spherical simulation with  $\gamma=2.5$}

\begin{figure}
\centerline{\includegraphics[width=0.5\textwidth]{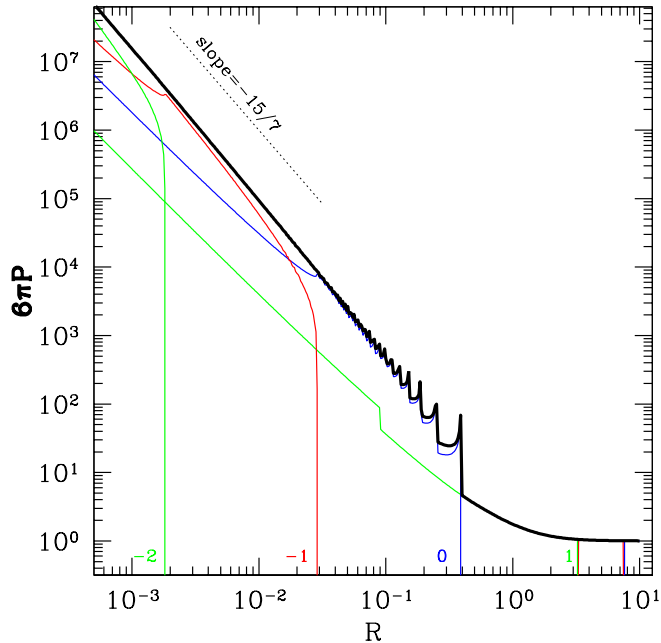}}
\caption{
Density Profiles (spherical symmetry, $\gamma=2.5$):\label{fig:sphere2gam25}
The black curve shows the self-similar density profile $P$, normalized
by the background density $1/6\pi$. At small 
$R$, it asymptotes to the frozen slope $-g_f=-15/7$.  The 
colored curves show ``shell profiles,'' i.e. the density contributed by
particles that initially lay within the same thin shell.  For example, 
the curve labelled $-2$ shows the density of particles that
have $10^{-3}<M_i<10^{-2}$, and subsequent curves show subsequent
decades in $M_i$.
}
\end{figure}

\begin{figure}
\centerline{\includegraphics[width=0.5\textwidth]{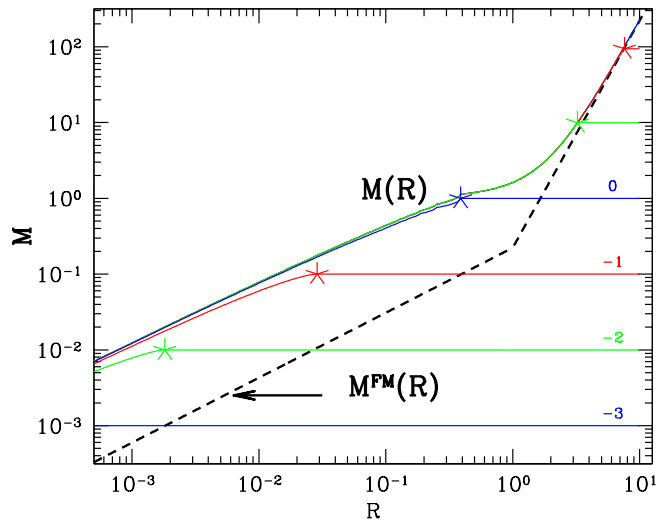}}
\caption{
Enclosed Mass Profiles  (spherical symmetry, $\gamma=2.5$): \label{fig:spheregam25}
 Each colored curve shows the mass profile from all shells
 interior to a given initial shell.  For example, the blue curve
 labelled 0 
shows the enclosed mass profile from all shells with 
$M_i<10^0$.
The upper envelope of the colored curves shows the enclosed mass
profile.  The dashed line is the mass profile in the frozen model.
The stars on each shell profile indicate the location of the shell's
current apoapse.
}
\end{figure}

\begin{figure}
\centerline{\includegraphics[width=0.5\textwidth]{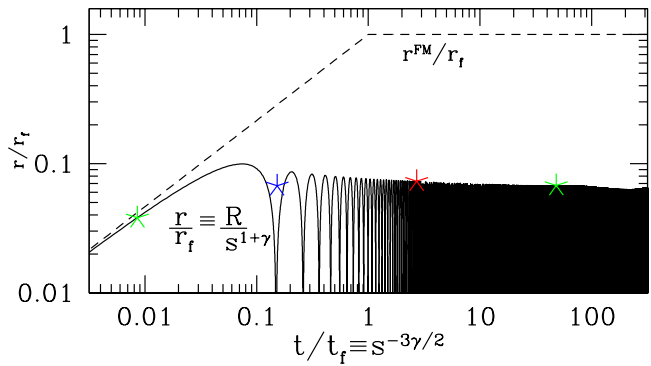}}
\caption{
Particle Trajectory (spherical symmetry, $\gamma=2.5$): \label{fig:trajgam25}
The solid curve shows a particle's proper radius versus time, 
where the radius is scaled relative to the frozen radius
and the time  relative to the frozen time (eq. [\ref{eq:rrf}]).
At late times, the apoapses are nearly constant.
The stars are equivalent to the ones in Figure \ref{fig:trajgam25}, converted
with equation (\ref{eq:mi}).  The dashed curve shows the 
trajectory in the frozen model.
}
\end{figure}

Figures \ref{fig:sphere2gam25}-\ref{fig:trajgam25} show results from 
 the spherical
solution
with $\gamma=2.5$.  
Figure \ref{fig:sphere2gam25} shows the total density profile ($P$) as a black line,
normalized to the background density ($=1/6\pi)$.
At small $R$, the slope is given by the frozen slope $-g_f=-15/7=-2.14$.
The colored lines show ``shell profiles,'' i.e. the density contributed by
particles that initially lay within the same thin shell, also normalized by $6\pi$.
   For example, the 
blue curve labelled 0 shows the density from all particles that have 
scaled initial mass
$10^{-1}<M_i<10^{0}$. 
  The red curve, which lies deeper in, comes
from a shell with smaller initial mass, $10^{-2}<M_i<10^{-1}$, etc.
The total density is the sum of the shell profiles.
It is apparent that the density at small $R$
is dominated by shells with small $M_i$, and the smaller
the $R$ the smaller the $M_i$ that contribute.

Figure \ref{fig:spheregam25} shows the enclosed mass profiles 
\be
M(R)\equiv 4\pi\int_0^R P(R')R'^2dR' \ .
\label{eq:m}
\ee
The upper envelope of the colored curves is the $M(R)$
that results from the total density profile shown in Figure \ref{fig:sphere2gam25}.
The individual colored curves show the enclosed mass profiles 
from all shells interior to a given shell.  For example, the blue curve labelled 0 
shows the enclosed mass profile from all shells with 
$M_i<10^0$, the red curve labelled -1 shows the profile from
all shells with $M_i<10^{-1}$, etc.  Note that 
the blue curve asymptotes at large $R$ to $M=10^0$, the red
to $M=10^{-1}$, etc.

Also shown in Figure \ref{fig:spheregam25} as a dashed line is the mass profile in 
the frozen model (see Fig. \ref{fig:frozen}).
The outer radius of each colored profile is marked with a star.  From the 
fact that the stars are parallel to the frozen profile at small $R$ indicates
that particles' apoapses do not continue to shrink after collapsing\footnote{
Although the break in the frozen model curve in Figure
\ref{fig:spheregam25} occurs at $R=1$, this location is arbitrary.
It can be shifted along the line $M=(2/9)R^3$ by requiring particles
to freeze at $r=kr_*$ with $k$ an arbitrary constant.}.

The  apoapse evolution may also be seen directly in Figure \ref{fig:trajgam25}, which
shows the time evolution of the proper radius of a single particle (and hence
of all particles).  The self-similar solution yields the function $R(s)$,
which we convert  to the function $r(t)$ (scaled relative to the particle's
$r_f$ and $t_f$) with the aid of equations (\ref{eq:rrf}).  
The particle's proper radius reaches turnaround at $t/t_f\sim 0.08$.  Thereafter, 
its apoapse remains roughly constant (although it does shrink very slightly at late times).
Also 
shown in Figure \ref{fig:trajgam25} are the stars from Figure \ref{fig:spheregam25},
where the co-ordinates $(R,M\equiv M_i)$ of each star in the latter figure
are converted to Figure \ref{fig:trajgam25} with the aid of equations
(\ref{eq:mi}) and
 (\ref{eq:rrf}); the stars clearly trace the particle's apoapse.

\subsection{Spherical Simulation with $\gamma=0.25$}

\label{sec:s2}

\begin{figure}
\centerline{\includegraphics[width=0.5\textwidth]{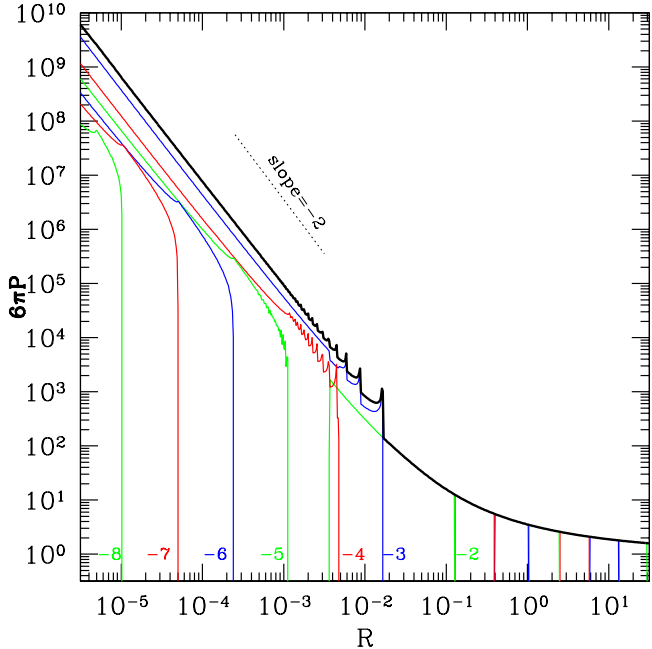}}
\caption{
Density Profiles (spherical symmetry, $\gamma=0.25$): \label{fig:sphere2}
The black curve shows the total density, and the colored curves
the shell profiles;  see caption of Fig. \ref{fig:sphere2gam25} for  detail.
The total density has logarithmic slope $-2$, much steeper than 
the frozen slope, because it is dominated by particles
with $10^{-4}<M_i<10^{-3}$ (the blue curve labelled -3).
}
\end{figure}

\begin{figure}
\centerline{\includegraphics[width=0.5\textwidth]{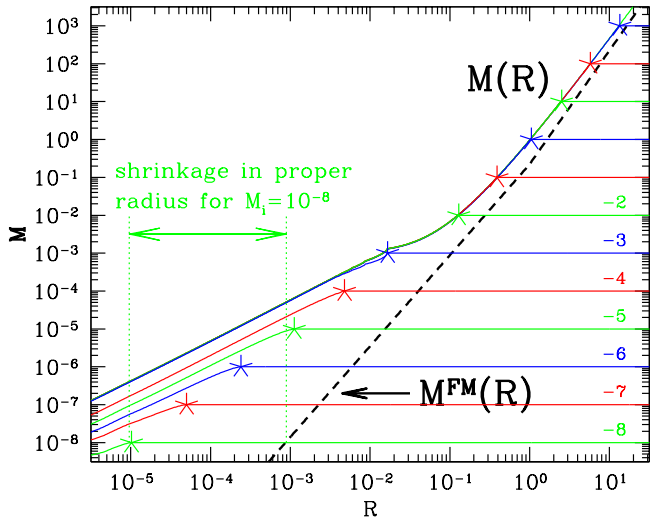}}
\caption{
Enclosed Mass Profiles (spherical symmetry, $\gamma=0.25$): \label{fig:sphere}
See caption of Figure \ref{fig:spheregam25} for description.  
}
\end{figure}

\begin{figure}
\centerline{\includegraphics[width=0.5\textwidth]{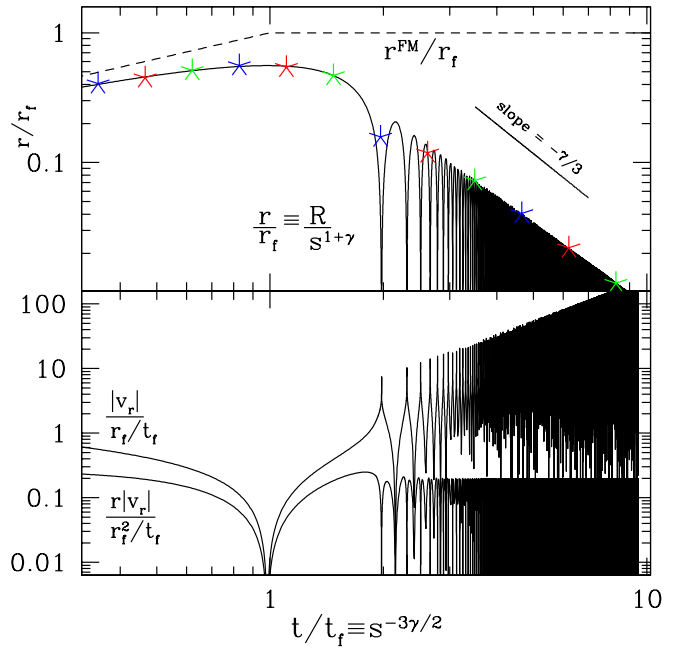}}
\caption{Particle Trajectory and Velocity \label{fig:traj}
(spherical symmetry, $\gamma=0.25$):  
The top panel shows a particle's proper radius versus time.
At late times, the apoapses shrink.
The bottom panel shows the radial velocity and the product
$r|v_r|$ relative to the frozen values.  From the fact
that the envelope of the $r|v_r|$ oscillations remains
constant, we infer that the radial adiabatic invariant
is very nearly constant.
}
\end{figure}


Figures \ref{fig:sphere2}-\ref{fig:traj} show results from a 
spherical simulation
with $\gamma=0.25$.  Throughout the rest of the paper, we will
frequently use $\gamma=0.25$ calculations to illustrate important
aspects of the self-similar solutions, in part because the effective
profiles of peaks of Gaussian random fields in CDM cosmologies have
comparably shallow interior slopes \citep[see][for more detail]{Paper2}.
Figure \ref{fig:sphere2} shows the total density
profile, as well as individual shell profiles.  At small $R$, 
the logarithmic slope of the total density is $-2$, 
which is much steeper than than 
the frozen slope $-3\gamma/(1+\gamma)=-0.6$.
From the shell profiles it is clear why the total density
profile is more concentrated than the frozen profile:
the blue curve labelled -3, which represents
particles with $10^{-4}< M_i< 10^{-3}$, shows that these particles
lay down a density profile with slope $-2$.
These particles dominate the total density profile
all the way to $R\rightarrow 0$.  Hence the total
density slope reflects theirs.
This behavior is in sharp contrast to that seen 
in the $\gamma>2$ simulation (Fig. \ref{fig:sphere2gam25}), 
where the density profile at increasingly small $R$ is dominated
by particles with increasingly small $M_i$ \citep{FG84}.

The reason that the individual shell profile has
 slope $-2$ may 
be understood qualitatively as follows.  
(See \cite{FG84} for the quantitative calculation.)
If we consider only particles from the same initial 
shell, 
  then the
mass they contribute within a sphere of radius 
$R$
  is proportional to
 $\delta t(R)$, 
  the amount of time
 the particles spend inside of that radius in the course of an orbit.
 Here, we assume that $R$ is smaller than their apoapse, and that the mass
 profile is averaged over a timescale of order the orbital time; 
 we also ignore the distinction between the scaled and unscaled
 $R$ because at late times the scaling factor changes negligibly
 over the course of a single orbit.
Since the interior gravitational
potential depends weakly  on $R$, 
the speed of a particle remains fairly constant as it plunges from 
its apoapse, through $R=0$,
to its next apoapse.   Therefore $M(R)\propto \delta t(R)\propto R$, 
and the resulting density profile is proportional to $R^{-2}$.

Figure \ref{fig:sphere} shows the enclosed mass profile, as
well as the enclosed mass profiles from cumulated shells.
It is clear from this figure as well that particles
with a narrow range of $M_i$ dominate the total profile
all the way to $R\rightarrow 0$. This figure
shows a second feature of spherical solutions with $\gamma<2$:  the
stars that denote the outermost radii of cumulated shells
are not parallel to the profile of the frozen model (the
dashed line in the figure).  This shows that the apoapses
of shells with increasingly small $M_i$ 
decrease
relative to the frozen profile--i.e., that a given shell's apoapses
shrink in time.

The top panel of Figure \ref{fig:traj} shows the shrinking of the apoapses directly.
The bottom panel shows the radial velocity, scaled appropriately, 
as well as its product with the radius.   
After the particle has collapsed, its radial adiabatic invariant is proportional to
 $\oint v_r dr$, integrated over an orbit.  
 From the fact that the envelope of oscillations
 of $r|v_r|$ remains constant, we infer that the adiabatic
 invariant is constant after collapse.

The constancy of the adiabatic invariant, together with the virial theorem, 
gives the rate at which the apoapse radius,
$r_a$, shrinks at late times \citep{FG84}.  The interior mass
profile scales linearly with radius:
$m(t,r)={\rm const}\times (r_*^3/t^2)(r/r_*)$, where the factors of $r_*$ and
$t$ are as required to convert from scaled to unscaled variables.
Therefore the virial theorem ($m(t,r_a)={\rm constant}\times v^2 r_a$,
where $v$ is a typical speed---e.g. the r.m.s. speed) together
with the constancy 
of the adiabatic invariant ($r_av=$constant)
 imply that the
amplitude of radial oscillations
shrinks in inverse proportion to the enclosed mass,
 \be
 r_a\propto 1/m(t,r_a) \ ,
 \label{eq:rm}
 \ee
  and hence
\be
r_a\propto t^{-({2\over 3\gamma}-{1\over 3})}  \ ,  \ \ \gamma<2 \ , 
\ee
\citep{FG84}.
The top panel of Figure \ref{fig:traj} confirms that $r_a\propto t^{-7/3}$ at
 late times.\footnote{More precisely, the slope in Figure \ref{fig:traj} is
-2.1, and not -7/3=-2.33.  The reason for this discrepancy is that the 
mass profile in Figure \ref{fig:sphere} actually has a slope of 1.1, not 1 
in this range of radii.}

\section{Axisymmetric Solutions}
\label{sec:axi}

For axisymmetric solutions, the linear
density field is prescribed to be $\propto f(\theta)R^{-\gamma}$ (eq. [\ref{eq:linP}]), 
i.e. $f$ is independent of
$\phi$ . This implies that particles can never acquire a $\phi$-component
to their velocities, and hence they must cross through the central axis
of symmetry repeatedly after collapse.

In spherically symmetric solutions
 an individual shell's density profile
has logarithmic slope -2 after collapse.  Therefore 
whenever $g_f<2$, the slope of an individual shell 
is steeper than that of the frozen slope, and hence recently 
collapsed shells dominate the interior density, driving
the interior total density to a slope of -2.

But in axisymmetric solutions, an individual shell's density
profile has logarithmic slope -1.
To see this, we consider the cylindrical co-ordinates of a particle
$(R_{cyl},\theta_{cyl},Z_{cyl})$.  Because of the axisymmetric assumption, 
the particle's $\theta_{cyl}\equiv \phi$ remains constant over the course of
its orbit, and only its $R_{cyl}$ and $Z_{cyl}$ change. 
As it traces out a trajectory in the $R_{cyl}-Z_{cyl}$ plane,
its speeds in the $R_{cyl}$ and $Z_{cyl}$ directions are nearly
constant (\S \ref{sec:s2}), and for a typical particle we may take
these two speeds to be comparable to each other.  
If we assume that the particle's orbit is chaotic, 
the amount of time it spends within a distance $R$ of the origin
(where $R$ is less than its apoapse) is proportional to the area
of that region in the $R_{cyl}-Z_{cyl}$ plane, i.e. 
 $\delta t(R)\propto R^2$.
Equivalently,
the amount of time a particle spends inside a sphere
of radius $R$ centered on the origin  also
scales as $\delta t(R)\propto R^2$.
 Since the mass enclosed within that sphere
is proportional to $\delta t$, it follows that $M\propto  R^2$, 
and hence that the density scales as $P\propto R^{-1}$.

Reasoning as before, we
 conclude that axisymmetric solutions should 
have interior density that scales as $P\propto R^{-g}$ , where
\be
g =
   \cases{
      g_f\equiv 3\gamma/(1+\gamma), & for $\gamma>1/2$\cr
     1, & for $\gamma<1/2$\cr 
   }
   \label{eq:gaxi}
\ee
\citet{Ryden93} calculated axisymmetric solutions for two 
values of $\gamma>1/2$---specifically, $\gamma=1$ and 2.  The resulting interior
density profiles have logarithmic slopes that are indeed close to 
$-g_f$, i.e. to -1.5 and -2, 
respectively, in agreement with equation (\ref{eq:gaxi}).

\begin{figure}
\centerline{\includegraphics[width=0.5\textwidth]{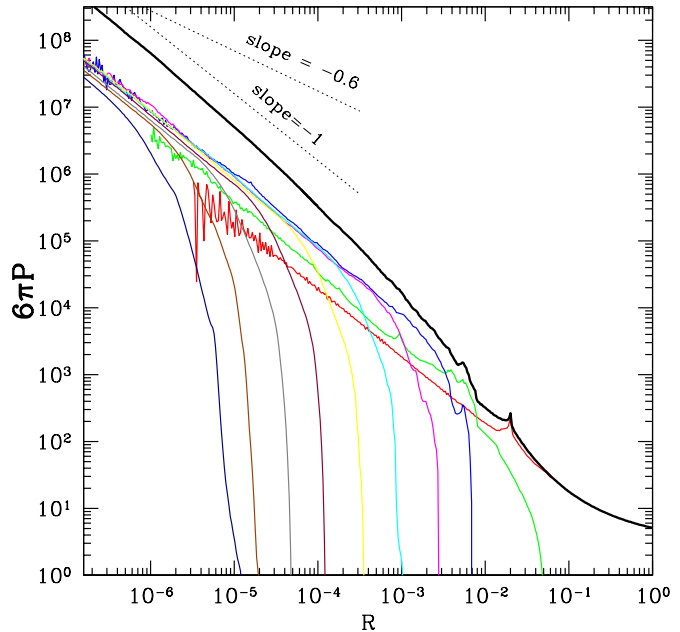}}
\caption{Axisymmetric Solution ($\gamma=0.25$, $e=0.1$, $p=0.1$):
 \label{fig:axi}
The black curve shows the total angle-averaged density profile.
The colored curves show the shell profiles.  
The frozen slope for this value of $\gamma$ is 
$-g_f=-0.6$.   Since the profile of an individual
shell has logarithmic slope -1, steeper than the
frozen slope, recently collapsed shells dominate
the interior total density, and drive its slope to -1.
 The largest shell profile (red) is
from particles with $M_i>10^{-3.4}$;  the next shell profile
(green) is from $10^{-4.4}<M_i<10^{-3.4}$, 
and subsequent profiles are from subsequent decades in $M_i$.
\newline
}
\end{figure}
Figure \ref{fig:axi} shows an axisymmetric simulation with $\gamma=0.25$
(and $e=p=0.1$; see Appendix \ref{sec:ep}).  
This simulation confirms that $g=1$ when $\gamma<1/2$.

Therefore the interior slope in axisymmetric collapse is
always $\leq -1$.  This is suggestive of NFW, where
the interior slope asymptotes to -1.
However, real halos are triaxial, not
axisymmetric.  As we show below, the slope in
a triaxial
halo rolls over to the frozen slope $-g_f$ even when $\gamma<1/2$. 
But this roll over can extend many decades in radius.

\section{Triaxial Solutions}
\label{sec:results}

In this section, we present fully three-dimensional
 self-similar solutions, and analyze them in detail.  
Each solution is  specified by the value of $\gamma$ 
and by the arbitrary angular function $f(\theta,\phi)$ in the linear density
field (eq. [\ref{eq:linP}]).
We choose the angular function
$f$ to be a sum of a monopole and a quadrupole field, 
\be
f(\theta,\phi)=1+a_{20}Y_{2,0}+a_{22}
\left(Y_{2,2}+Y_{2,-2}\right) \ ,
\ee
where the $Y_{l,m}(\theta,\phi)$ are spherical harmonics, 
and $a_{20}$ and $a_{22}$ are arbitrary
constants.  We parameterize these constants
in terms of the ellipticity and prolateness ($e$ and $p$)
of the tidal tensor. See equations \ref{eq:a20}-\ref{eq:a22}
in Appendix \ref{sec:ep}. 
Unless explicitly stated otherwise, we always set 
 $p=0$, which yields
a triaxial ellipsoid that has middle axis equal
to the average of the large and small axes.  
Note that $e=0$ corresponds to the spherically symmetric case, 
and the ellipticity increases with increasing $e$.
In general, one may choose $f$ to be any sum of spherical 
harmonics.  
Preliminary investigations
indicate that typically the monopolar and quadrupolar terms
are the most important, but this should be investigated
in more detail in the future.

Our 3D simulations
typically include
spherical harmonics up to
 $l_{\rm max}=28$ and
 ${\cal O}(10^7)$ particles. The radial
 range covered is up to 16 decades in $R$.

\subsection{Density Profiles}

\begin{figure}
\centerline{\includegraphics[width=0.5\textwidth]{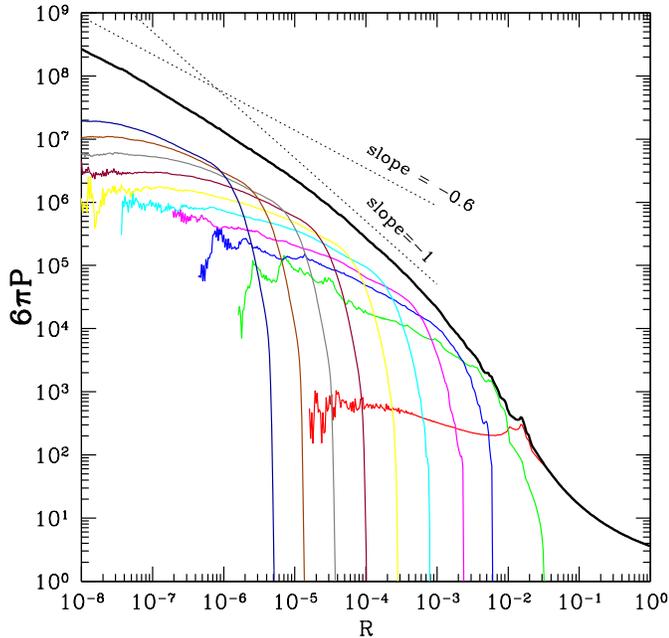}}
\caption{
Triaxial Solution ($\gamma=0.25$, $e=0.1$). \label{fig:rhoplot}
The black curve shows the total angle-averaged density profile, which
rolls towards the frozen slope $-g_f=-0.6$ over many decades
in $R$.
The colored curves show the shell profiles. 
The largest shell profile (red) is
from particles with $M_i>10^{-3.4}$; the next smaller one
(green) is from $10^{-4.4}<M_i<10^{-3.4}$, 
and subsequent profiles are from subsequent decades in $M_i$.
}
\end{figure}

\begin{figure}
\centerline{\includegraphics[width=0.5\textwidth]{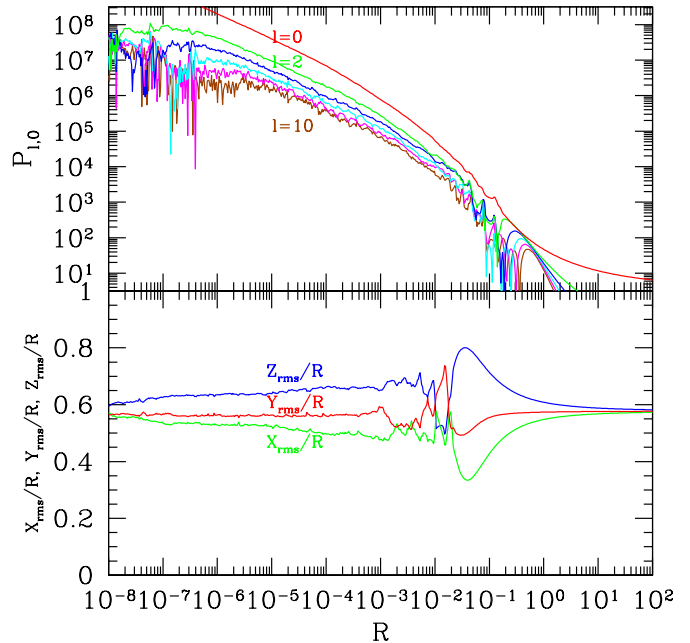}}
\caption{
Higher Multipole Moments in a Triaxial Solution 
($\gamma=0.25$, $e=0.1$). \label{fig:multipole}
The top panel shows selected density multipoles, 
specifically $P_{l,m}$ with $m=0$, and $l\leq 10$.
The bottom panel shows the mass weighted
co-ordinates at each $R$, which are directly
related to the quadrupole ($l=2$) moments; for example, 
$Z_{\rm mrs}^2/R^2\equiv 1/3+2/(3\sqrt{5})P_{2,0}/P_{0,0}$.
}
\end{figure}

Figure \ref{fig:rhoplot} shows the angle-averaged density 
profiles for the solution with $\gamma=0.25$ and $e=0.1$.
Comparing with the spherically symmetric case with 
the same $\gamma$ (Fig. \ref{fig:sphere2})
and with the axisymmetric case (Fig. \ref{fig:axi}),
shows that  the logarithmic slope of $P$ is much flatter
for the 3D case, and nearly reaches the frozen slope $-g_f=-0.6$ at very small $R$.
The spherically symmetric case has $g=2$ because its interior
density field is dominated
by particles that have recently collapsed and themselves
lay down a shell profile with logarithmic slope -2.  By contrast, in the
triaxial case the recently collapsed shells have flatter tails. 
Physically, this is because non-radial motions prevent recently
collapsed particles from penetrating the origin, whereas in the
spherical case all particles are forced to go through the origin
every orbital time \citep{WhiteZaritsky92,Nusser01}.
From Figure \ref{fig:rhoplot}, it is apparent that the density
field at increasingly small $R$ is dominated by particles with 
increasingly small $M_i$.  This is in contrast to the  spherical $\gamma=0.25$
case (Fig. \ref{fig:sphere2}) and to the axisymmetric $\gamma=0.25$ case
(Fig. \ref{fig:axi}), but similar to the spherical $\gamma=2.5$ case
(Fig. \ref{fig:sphere2gam25}).

A surprising aspect of Figure \ref{fig:rhoplot} is how 
many decades in $R$ it takes the total density
to roll over towards the frozen slope.
  Indeed, the slope has not yet reached
$-g_f$ 
even at $R\sim 10^{-7}$, which is 
$\gtrsim 10^5$ times smaller than the virial radius.
 For 
$R$ smaller than that, it appears to reach $-g_f$, but 
 the simulation has likely not converged at such 
small $R$.
We shall discuss the reasons for the gradual roll-over of the
slope below.

Figure \ref{fig:multipole} shows some higher order multipoles in 
the same solution as in Figure \ref{fig:rhoplot}, quantifying the
deviation from spherical symmetry. 
At a given $R$, the mass weighted r.m.s. co-ordinates
differ from each other by around $30\%$.

\begin{figure}
\centerline{\includegraphics[width=0.5\textwidth]{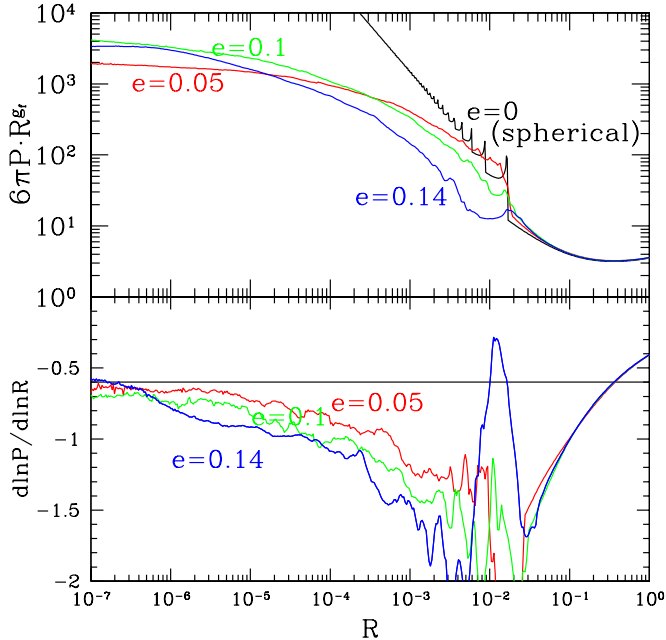}}
\caption{
Density Profiles ($\gamma=0.25$, various
values of $e$): \label{fig:rhos}
The density profiles (top panel) have been compensated 
by
 $R^{g_f}$ to highlight their approach to the frozen slope at
 small $R$.
The bottom panel shows the logarithmic slopes of the densities
in the top panel. The frozen slope is $-g_f=-0.6$, indicated with a horizontal line.
}
\end{figure}
\begin{figure}
\centerline{\includegraphics[width=0.5\textwidth]{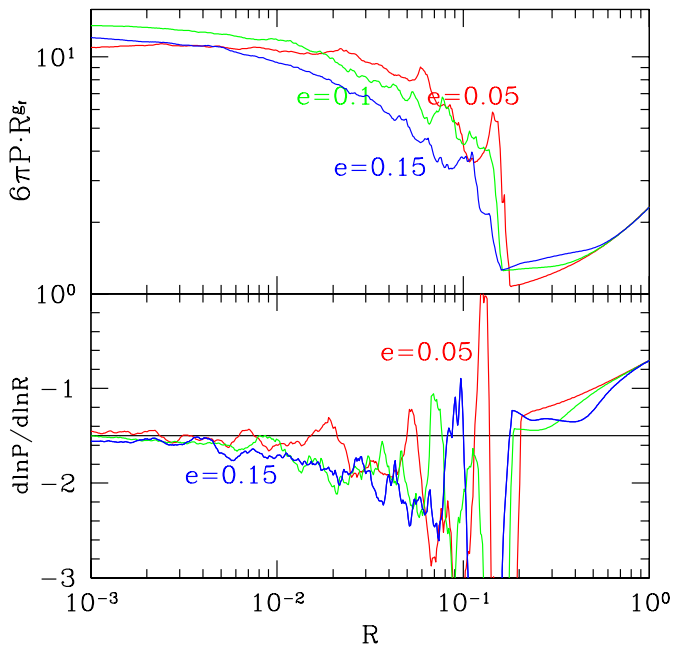}}
\caption{
\label{fig:rhosgam1}
Density Profiles
 ($\gamma=1$, various
values of $e$):   similar to Fig. \ref{fig:rhos}.  The frozen slope
is -1.5 at this $\gamma$.
}
\end{figure}
\begin{figure}
\centerline{\includegraphics[width=0.5\textwidth]{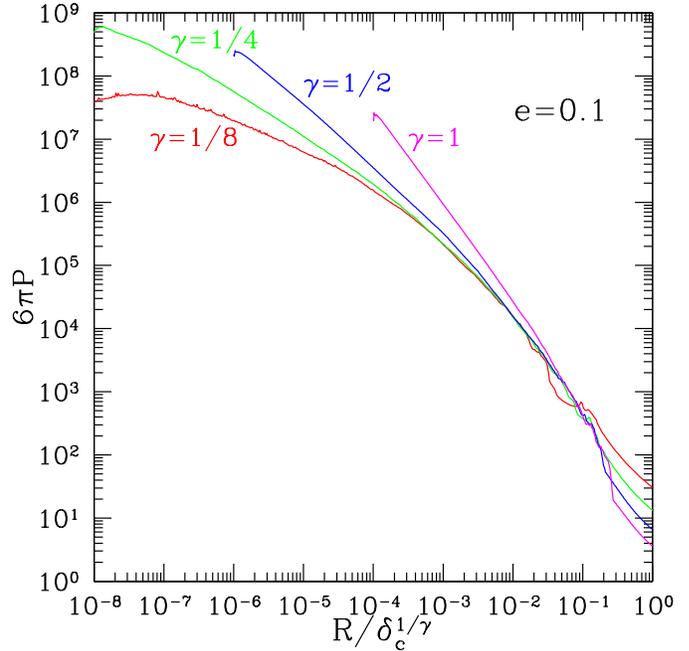}}
\caption{
\label{fig:rhogamma}
Density Profiles ($e=0.1$, various values of $\gamma$):
Simulations with smaller $\gamma$ reach smaller
asymptotic slopes.
The $x-$axis has been rescaled by the $\delta_c^{1/\gamma}$
to roughly line up the virial radii of the various simulations,
where $\delta_c\equiv 1.686$ is the critical density for top-hat
collapse
}
\end{figure}
Figures \ref{fig:rhos}-\ref{fig:rhosgam1}  show the spherically-averaged density 
profiles for a number of solutions with various values
of $e$ (and $p=0$),  and 
with $\gamma=0.25$ and 1, respectively.
In the top panels, the profiles have been multiplied 
 by $R^{g_f}$ to show more clearly where they 
reach the frozen slope. The bottom panels show the
logarithmic density slopes.
For the triaxial  simulations, increasing $e$ at fixed $\gamma$
tends to makes the profiles steeper, i.e. more discrepant with 
the frozen slope, and extends the range in $R$ over which
the slope rolls over towards the frozen slope.\footnote{
  See  \S \ref{sec:toy} for a possible explanation.
  This  trend has potential implications for the structure of cosmological
  halos.  Since we find that increasing $e$ has the effect of slowing
  the roll-over in slope from steep to shallow, we might expect that
  high-mass halos, which form from relatively spherical peaks, will
  have a faster roll-over than low-mass halos, which tend to form from
  peaks of high ellipticity \citep{BBKS}.  There are possible hints of
  such behavior in $\Lambda$CDM simulations \citep{Gao08}.}
It appears that all profiles do reach the frozen slope
at small $R$.
At larger $R$, increasing $e$
tends to spread out the outermost caustics, 
and
pushes the rise in density below the caustic region
 to smaller radii.

Figure \ref{fig:rhogamma} shows a set of solutions with $e=0.1$
and varying $\gamma$.
  Note that the $x-$axis 
has been rescaled to roughly line up the virial radii. 
Decreasing $\gamma$ tends to decrease the 
asymptotic slope at small $R$.

\subsection{Caustics}

\begin{figure}
\centerline{\includegraphics[width=0.26\textwidth]{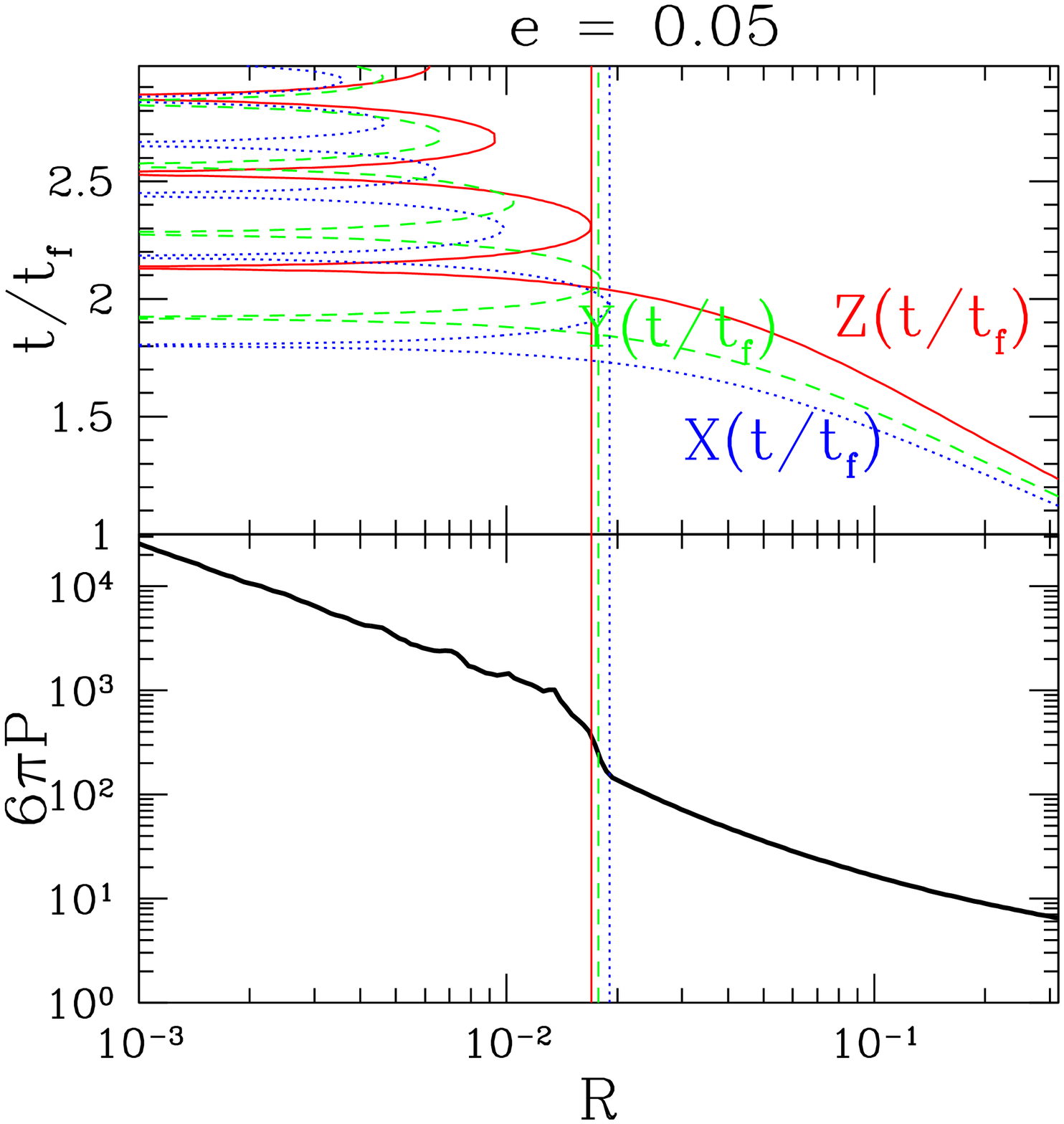}
\includegraphics[width=0.26\textwidth]{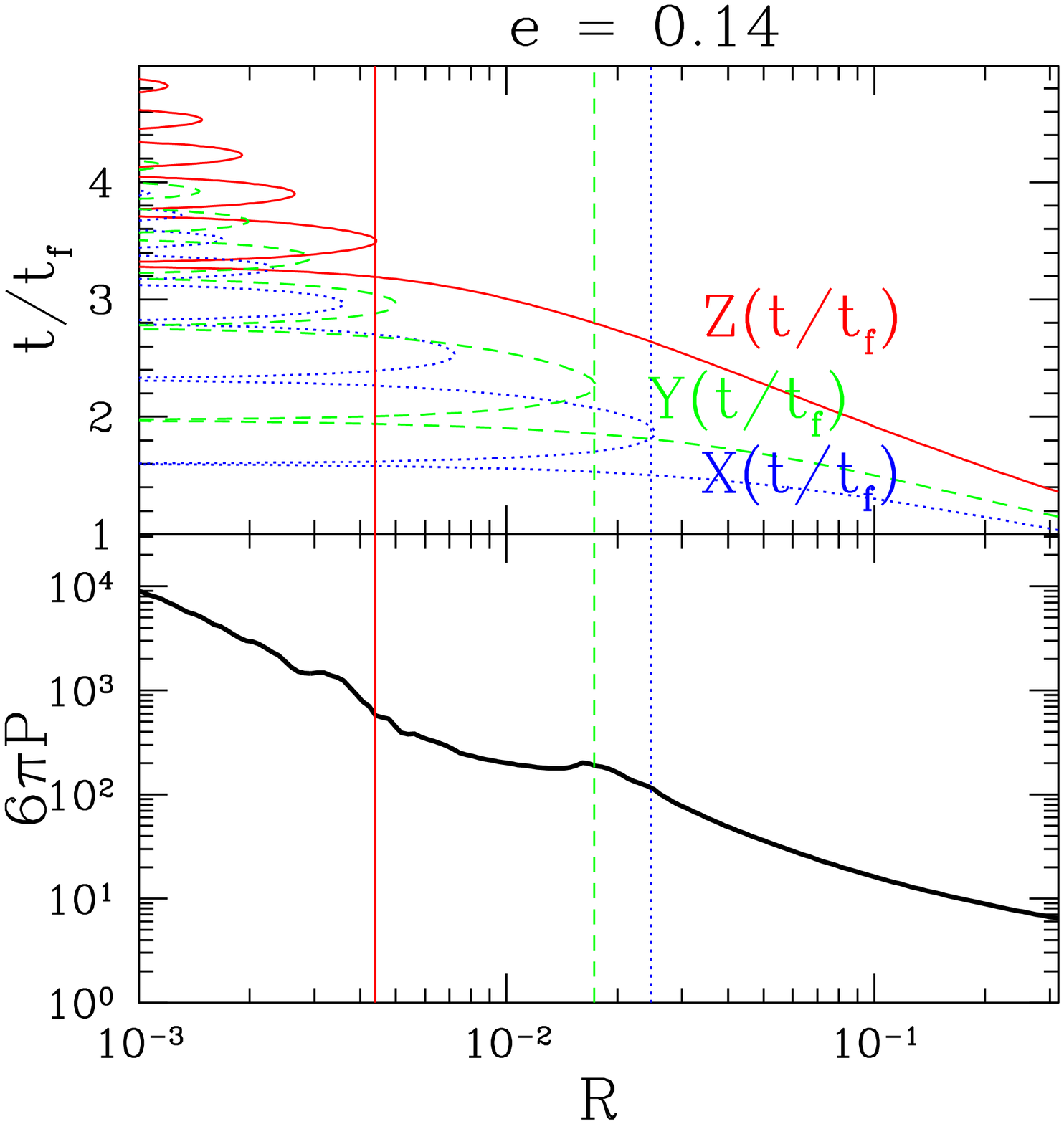}}
\caption{
Caustics  \label{fig:caustics0.05}
($\gamma=0.25$):
For the $e=0.05$ case (left), the top panel 
shows the {\it scaled} co-ordinates of the three axis particles.  
The bottom
panel shows the density profile.  Vertical lines
are drawn where the three axis particles reach
their first respective
apoapses, in scaled co-ordinates.
 These locations  correspond
 to the sharp rise in the density
profile in the bottom panel.
For the $e=0.14$ case (right), 
the time at which the long-axis particle ($Z$)
hits its first scaled apoapse is significantly delayed relative to 
the short-axis particle ($X$).  This is reflected
in the density profile (bottom panel) by an extended flat
region, and then a rise in density below $R\sim 0.004$, 
which is
where the long-axis particle hits its first scaled apoapse.
}
\end{figure}

The density profiles of spherically symmetric solutions exhibit
sharp spikes that are especially prominent at large radii (e.g., Fig. 
\ref{fig:sphere2gam25}).  These  caustics
occur near a particle's apoapse, where its speed vanishes
 \citep{FG84,Bert85}.
For triaxial halos,  caustics  are largely washed
out in the spherically averaged density profiles (Fig. \ref{fig:rhos}),
because particles with different initial angular positions reach their
apoapses at different radii and times.

In Figure \ref{fig:caustics0.05} we examine
caustics in more detail for two simulations, one
with $e=0.05$ and the other with  $e=0.14$.
The bottom panel of the $e=0.05$ case
 shows a 
zoom-in of the density profile.
  Although most of the caustics
present in the spherical run are washed out, 
similar to the results of \citet{VMW10},
there is a clear 
rise in density at $R\sim 0.02$, which is where the
spherical run exhibits its first caustic (Fig. \ref{fig:rhos}).
The top panel shows
the trajectories of the three axis particles;
these are the particles that lie along the principal
axes of the  linear density field.
For example, for the particle on the $z$-axis  (which we
take to be the long axis), the self-similar solution determines
$Z(s)$, i.e. the $z$-component of $\bld{R}(s)$.  In the top  panel,
 the solid red line shows this $Z$ as the abscissa, and 
the time as the ordinate, where
the time is converted from $s$ to $t/t_f$ via equation  (\ref{eq:rrf}).
Similarly, the blue and green curves show the particles
that lie along the two shorter principal axes.
At early 
times, towards the bottom of the panel, all three particles
expand with the Hubble flow, and their scaled radii
$X,Y,Z$ decrease together.  Then, at $t/t_f\sim 1.8$ the
particle along the short axis collapses and crosses
through $X=0$.  
We emphasize that Figure \ref{fig:caustics0.05} depicts
{\it scaled} (i.e.\ self-similar) co-ordinates.  When $X$ crosses
through 0, the unscaled $x$ has already been through turnaround.
This is not evident in the figure because $X$ decreases uniformly
until it reaches the origin.  As a result, the time
and location of turnaround
have only a small influence on the density profile.
At $t/t_f\sim 2$, the short axis particle reaches its first 
scaled apoapse.  This is close to when the unscaled $x$ reaches
its first apoapse after turnaround.
The long axis particle, labelled $Z$, reaches its
first apoapse in scaled co-ordinates shortly after the
short axis particle, and at nearly the same scaled radius.
We also plot vertical lines at the locations of the first
(scaled) apoapses, 
and extend them to the bottom panel, showing that the sharp
rise in density is caused by the corresponding caustics.

The right half of the figure shows a similar plot, but with $e$ 
 increased to $0.14$.  
In the top panel, one sees that the first scaled apoapse of the long-axis
particle  occurs significantly after that of the
short-axis particle.
Because of this, the former is significantly
{\it smaller} than the latter,
a somewhat counterintuitive result.  This behavior is reflected
in the density profile shown in the lower panel.  At the location
of the caustic of the short-axis particle, there is a modest 
rise in density.    The density does not change significantly until
reaching a much smaller radius ($R\sim 0.005$), close
to where the long-axis particle reaches its first scaled 
apoapse.

\subsection{Adiabatic Shrinking}
\label{sec:adshrink}

\begin{figure}
\centerline{\includegraphics[width=0.5\textwidth]{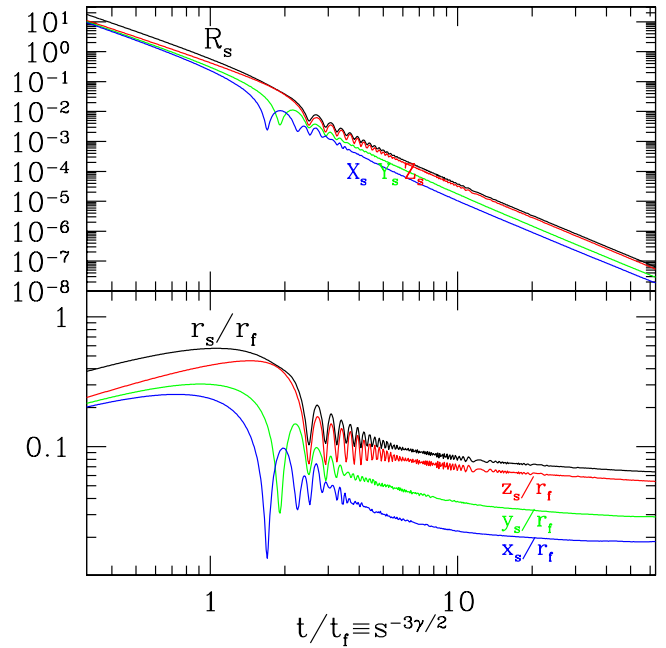}}
\caption{
Root-mean-square co-ordinates of the set of particles that
initially lie in the same spherical shell \label{fig:sxyz}
($\gamma=0.25$ and $e=0.1$).   The top panel 
shows the scaled r.m.s. co-ordinates, and the bottom panel
the unscaled (i.e. proper) ones. 
There is much less shrinking at late
times than in the spherical case (Fig. \ref{fig:traj}).
}
\end{figure}

Figure \ref{fig:sxyz} shows the time evolution of
 an initially thin spherical shell of particles
in the $\gamma=0.25$, 
$e=0.1$ solution.
In the top panel, we 
plot
 the r.m.s. value of $R$ for all particles that have the same $s$.
We also show the r.m.s. values of the $X$, $Y$, and $Z$ co-ordinates
for these particles. 
In the bottom panel, we convert the scaled co-ordinates of the 
top panel to unscaled ones (via eqs. [\ref{eq:rscal}] and [\ref{eq:rrf}]).
At early times, the r.m.s. co-ordinates  expand with
the Hubble flow, until first the small axis ($x_s$) 
turns around and 
collapses, then the middle, 
and finally the long axis.  At late times the axes continue to shrink, but
by a much smaller amount 
 than in the spherical simulation
 (Fig. \ref{fig:traj}).  
 Eventually, the axes become nearly constant.
 This explains why the slope of the density profile reaches
 the frozen slope at very small radii (see point 2 above \S \ref{sec:rftf}).
 Note also that at late times,  the
 r.m.s. co-ordinates are in the ratio $3.1:1.6:1$.

\subsubsection{Effect of shrinking on density profile}

\begin{figure}
\centerline{\includegraphics[width=0.5\textwidth]{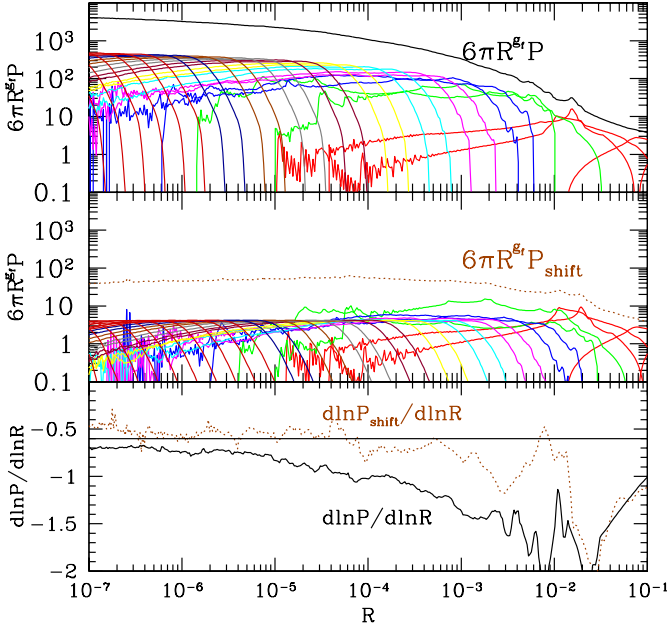}}
\caption{
Effect of shrinking on density profile
 ($\gamma=0.25$, $e=0.1$): \label{fig:noshrink}
 The top panel shows the shell densities and total density,
 multiplied by the frozen profile $R^{g_f}$.
 The middle panel takes the shell profiles from the top panel, 
 and ``unshrinks'' them;  see text for details.  The brown
 dotted line shows the sum of the shifted profiles.
The profiles in the middle panel have also been compensated
by $R^{g_f}$.  The bottom panel shows the logarithmic slope
of the total densities in the top two panels, showing
that the shifted profile is much closer to the frozen
one, with  logarithmic slope much closer to $-g_f=-0.6$.
}
\end{figure}
Surprisingly, the small amount of shrinking at late
times has a marked effect on the slope of the density
profile.  Figure \ref{fig:noshrink} illustrates this.
The top panel of Figure \ref{fig:noshrink} 
shows the density and shell profiles of the $e=0.1$, $\gamma=0.25$
solution, compensated by the frozen slope.
Note that twice as many shells have been plotted in this panel 
as in
 Figure \ref{fig:rhoplot}, each shell being half as wide logarithmically.
In the middle panel, we spatially expand each shell profile from the top panel,
 conserving mass,
 such that the r.m.s. unscaled radii do not
shrink;
 i.e.,
 we set $P_{\rm shifted\ shell}(R)=c^3P_{\rm shell}(cR)$, where
 the factor $c$ was chosen  to make the
   r.m.s. unscaled radius of each shifted shell
 equal to $r_s=0.5r_f$ (rather than declining in time, as seen
 in Fig. \ref{fig:sxyz}). The brown dotted line in the middle
 panel shows the sum of these shifted shell profiles.
The bottom panel of the figure shows the logarithmic derivatives
of the unshifted and shifted densities,
showing that the latter
is much closer to the frozen profile.
This figure illustrates that the late-time shrinking of the r.m.s.
radii plays a large role in causing the density profile to 
deviate from the frozen profile.

To see why the   small amount of shrinking  shown in Figure
\ref{fig:sxyz} leads to such a significant change in the
density profile, we construct a simple model, 
which is a slight extension of the frozen model
(\S \ref{sec:frozen}).  
After each particle crosses $r_*$, we now allow
its proper radius to decrease in time
as a power-law, i.e.
we set
\be
r= r_f\left({t\over t_f}\right)^{-\beta}  ,\  {\rm for\ } t>t_f
\ee
and ask what the resulting density profile would be.
Dividing through by $r_*$ and using
equations (\ref{eq:mi}) and (\ref{eq:rrf}), we find
$M_i={2\over 9}R^{3/( 1+\gamma+3\beta\gamma/2)}$.
Since in this model we ignore the post-collapse orbital
motion (i.e. we take the shell profiles to be delta functions), 
we have $M_i=M$, implying that $P\propto R^{-g}$ at small $R$,
where
\be
g=3\gamma{1+3\beta/2\over 1+\gamma+3\beta\gamma/2}\approx
g_f \left(1+{3\beta/2\over 1+\gamma}\right) \ ;
\ee
the approximation holds when $\beta\ll 1$.
So if $\beta\sim 0.1$, it would lead to a $\sim 15\%$ deviation in the
slope from the frozen slope.
For example Figure \ref{fig:sxyz} implies that at
 time $t/t_f=30$
 (when $R_s\simeq 10^{-6}$) that $\beta\simeq 0.13$, and hence
 the above equation yields
 $g\simeq 0.7$.  
Comparing with the lower panel of
 Figure \ref{fig:noshrink}, we see
that indeed the black line is nearly equal to $-0.7$ at
$R\simeq 10^{-6}$.

\subsubsection{Effect of density profile on shrinking}

\begin{figure}
\centerline{\includegraphics[width=0.5\textwidth]{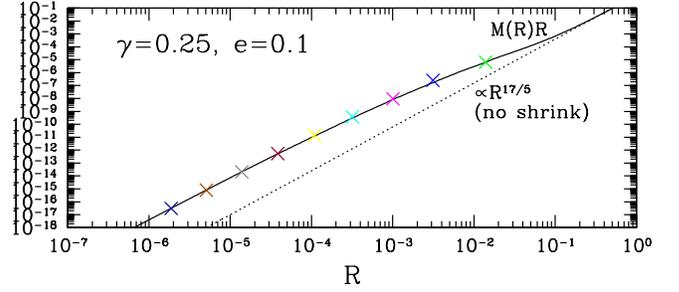}}
\centerline{\includegraphics[width=0.5\textwidth]{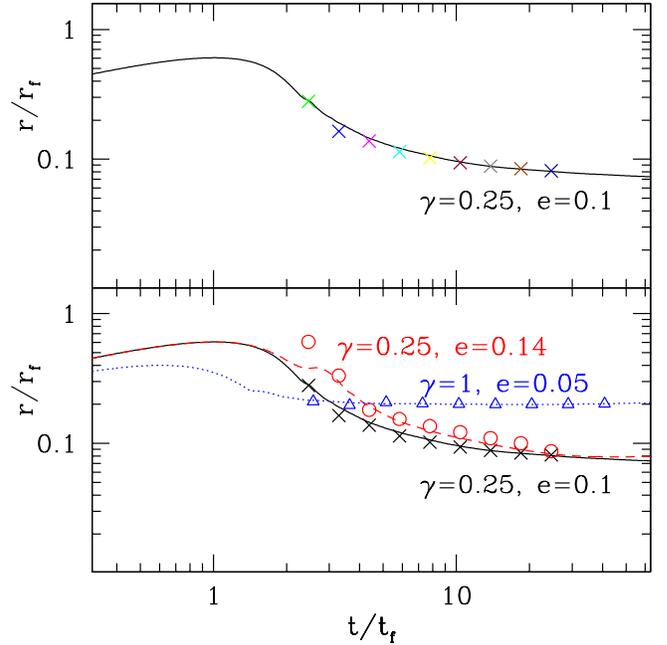}}
\caption{
Effect of density profile on shrinking: \label{fig:shrink}
These plots show that shells shrink in inverse proportion 
to the average enclosed mass (eq. [\ref{eq:mr2}]).
See text for details. The 
order-unity constant $k$ appearing in equation (\ref{eq:mr2})
was chosen to be 0.8 for the $\gamma=0.25$ simulations, 
and $0.4$ for the $\gamma=1$ simulation.
}
\end{figure}

In spherically symmetric solutions, the apoapse of a particle
shrinks
in inverse
proportion to the enclosed mass (eq. [\ref{eq:rm}]).
We show here that this is nearly true in 3D as well, 
even though the
radial adiabatic invariant need not be conserved in the absence of
spherical symmetry.
To show this, we note that 
if the 
radius shrinks in inverse proportion to the enclosed mass, then
\be
m(r)r={\rm const}=km_i r_f \ ,  \label{eq:mr2}
\ee
where we write the constant as a product of three factors:  
the frozen radius $r_f$ (eq. [\ref{eq:rrf}]), 
the initial enclosed mass (see above eq. [\ref{eq:mi}]), 
and an order unity
constant $k$.
Dividing through by $r_*^4/t^2$ yields the scaled equation
\be
M(R)R={2\over 9}ks^{4+\gamma} \ . \label{eq:mrs}
\ee
Hence the total enclosed mass profile $M(R)$ predicts
how the radius of a shell  decreases with decreasing 
$s$.  
The solid line in the top panel of Figure \ref{fig:shrink} shows
 the $M(R)R$
profile for the $\gamma=0.25$, $e=0.1$ simulation, 
where $M(R)$ is the angle-averaged enclosed mass
profile (eq. [\ref{eq:m}]) that comes from the density profile 
shown in Figure \ref{fig:rhoplot}.
The colored points show, for each of the shell profiles in 
Figure \ref{fig:rhoplot}, the r.m.s. value of $R$ as the abscissa, 
and $(2/9)ks^{4+\gamma}$ as the ordinate, where we select
$k=0.8$ to fit the $M(R)R$ curve.  From the fact that
the points lie on top of the curve shows that 
equation
(\ref{eq:mrs}) (and hence eq. [\ref{eq:mr2}]) is a good prescription for how 
the r.m.s. radii of the shells shrink in time.  By contrast, 
if the shells did not shrink (i.e. if $r=k'r_f$), they would have their
r.m.s. $R_s\propto s^{1+\gamma}$, which would give the dotted
line in Figure \ref{fig:shrink}, with slope $(4+\gamma)/(1+\gamma)=17/5$.
The middle panel of Figure \ref{fig:shrink} shows the same data
as that in the top panel, but converted to unscaled radius and time, i.e. 
the $M(R)R$ profile has been converted to $s(R)$ via equation 
(\ref{eq:mrs}), and then the $R$ and $s$ have been converted
to $r$ and $t$ via equation (\ref{eq:rrf}).
The points show the r.m.s. radii of shells, and are identical
to  the black curve of Figure \ref{fig:sxyz}, lower panel, except
that here we average over a wider range in $s$.
This middle  panel shows directly that the shrinking is due
to the increase of enclosed mass.  The bottom panel of 
Figure \ref{fig:shrink} repeats the data in the middle panel, and
also shows two other simulations, showing that in all cases
the shrinking is described quite well by 
equation (\ref{eq:mr2}).

\begin{figure}
\centerline{\includegraphics[width=0.5\textwidth]{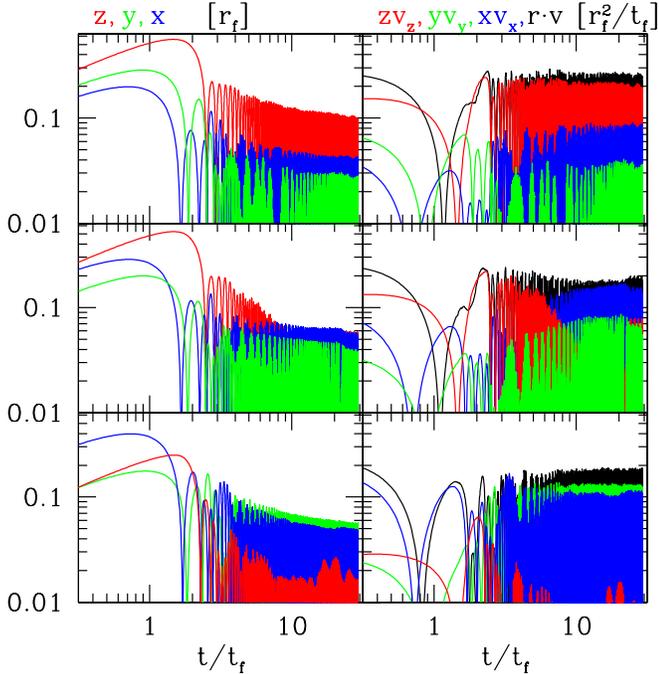}}
\caption{
\label{fig:particles}
Trajectories of three
randomly selected particles ($\gamma=0.25$, $e=0.1$):
For each particle, the left column shows its co-ordinates, 
normalized to the frozen radius $r_f$; 
the right column shows the product of its co-ordinates with
the corresponding speed, appropriately normalized, as well
as the sum of the three ($\bld{r\cdot v}$).  
The envelope of $\bld{r \cdot v}$ is nearly constant, 
even at rather early times, indicating that the radial adiabatic
invariant is nearly constant for each particle.
}
\end{figure}
The fact that the radius decreases in inverse proportion to the
enclosed massed is surprising, since for nonspherical potentials
the radial adiabatic invariant need not be constant.
And  halos
deviate significantly from spherical symmetry: at fixed
$R$, the mass-weighted co-ordinates differ from 
each other by $\sim 30\%$ for the 
$e=0.1$ case (Fig. \ref{fig:multipole}), and
by $\sim 50\%$ for the $e=0.14$ case.
Figure \ref{fig:particles} exhibits an even more
striking result:  that the radial adiabatic invariant of 
individual particles is nearly constant.
For that figure, we randomly selected three particles
in the $\gamma=0.25$, $e=0.1$ simulation.  
(All other particles  that we examined display similar behavior.)
The three left panels show the unscaled co-ordinates of
those three particles.  And the right panels show
the corresponding values of the co-ordinates multiplied
by the respective co-ordinate speed.  From the fact
that the product $\bld{r\cdot v}=rv_r$ has a nearly constant
envelope, we conclude that the radial adiabatic
invariant $\propto\oint v_rdr$ is very nearly 
constant.

\subsection{Shell Profiles}
\label{sec:shellprof}

\begin{figure}
\centerline{\includegraphics[width=0.5\textwidth]{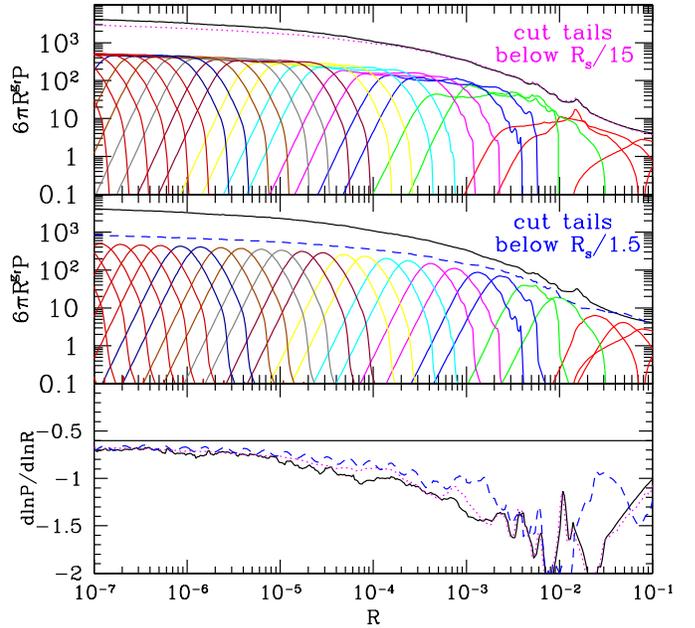}}
\caption{\label{fig:notails}
Effect of the shell tails on the density profile ($\gamma=0.25$, $e=0.1$):
The top panel shows the shell profiles from 
Figure \ref{fig:noshrink}, but truncated below 
$R_s/15$, where $R_s$ is the r.m.s. radius of each shell.
The dotted magenta line is the sum of the
truncated profiles, which deviates slightly from the
untruncated  profile (black line).  
All profiles are compensated by $R^{g_f}$.
The middle panel is the same, except truncated below
$R_s/1.5$.
The bottom panel shows the logarithmic derivatives of the
total density profiles from the top two panels.
}
\end{figure}

The tails of the shell profiles 
extend many orders of magnitude in radius (Fig. \ref{fig:noshrink}).
  How important
are these tails for the shape of the total density profile?
The top panel of Figure \ref{fig:notails} shows
the density profile that results from truncating the tails
of the shell profiles below $R_s/15$, where
$R_s$ is the r.m.s. radius of a shell.
Specifically, for each shell profile shown in Figure \ref{fig:noshrink}, 
we compute the enclosed mass profile, and then 
multiply it by $x^5/(1+x^5)$, where $x\equiv R_s/15$.
The resulting density profiles are shown, 
as well as their sum, which is plotted as a magenta dotted line.
The black line is the total (untruncated) profile.  The middle panel
is similar, except the shell profiles are truncated below $R_s/1.5$.
The bottom panel
shows the logarithmic derivatives of the total density profiles
from the top two panels.
We conclude that the deep tails of the shell profiles
($\lesssim R_s/15$), 
have little effect on the total density.
But the moderately deep tails
have a noticeable impact,
particularly for $R\gtrsim 10^{-4}$, 
where the adiabatic shrinking is insufficient to account for the total
deviation from the frozen slope (Fig. \ref{fig:noshrink}).

Therefore both effects---shell tails and the adiabatic shrinkage---influence the 
total density profile.  Of course, our distinction between these
two effects is somewhat artificial.  It is the shell tails that increase
the enclosed mass, which in turn causes adiabatic shrinkage.
In the absence of shell tails there could be no adiabatic shrinking, because
the mass enclosed by any shell would not change in time.
We construct a toy model that incorporates both effects self-consistently
below (\S \ref{sec:toy}).

\begin{figure}
\centerline{\includegraphics[width=0.5\textwidth]{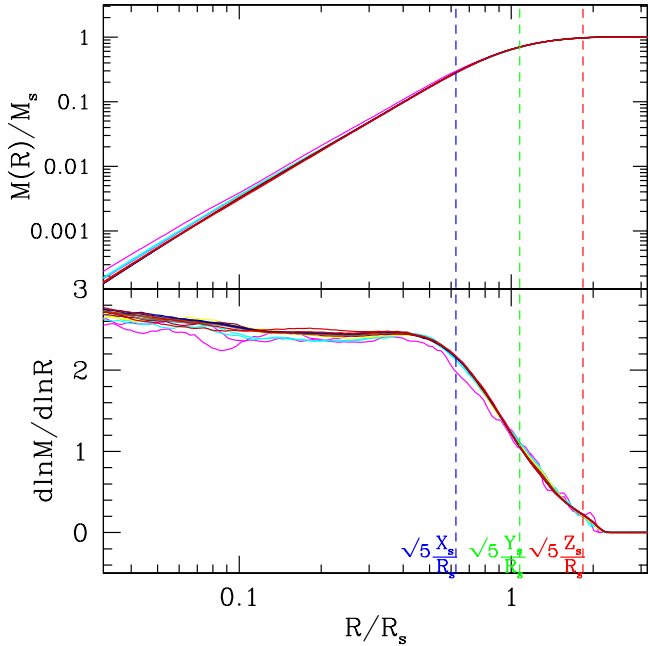}}
\caption{\label{fig:shells}
Normalized Shell Profiles ($\gamma=0.25$, $e=0.1$):
The top panel shows 16 shell profiles from 
Figure \ref{fig:noshrink}'s top panel, 
but plotted here as enclosed mass rather than density, 
and with the mass and radius normalized by 
$M_s$ and $R_s$ (mass in the shell and r.m.s. radius of shell).
From the fact that all of the shell profiles lie nearly 
on top of each other, we conclude that there is little evolution
of the shell profile.
The bottom panel shows the logarithmic derivatives of the
profiles in the top panel.  The vertical lines are the  r.m.s.
co-ordinates relative to $R_s$ multiplied by 
$\sqrt{5}$, as would be applicable for a constant
density ellipsoid.
}
\end{figure}

Figure \ref{fig:shells} shows many of the shells from Figure \ref{fig:noshrink}, 
but replotted
with a common normalization.
Specifically, we take the 16 shells with $4.3\leq t/t_f\leq 36$, where the
lower limit is the first magenta shell at large $R$, and the upper limit
is, continuing to smaller $R$, the second red shell. For each shell, 
we plot in the top panel of Figure \ref{fig:shells} its enclosed mass
normalized to the total mass in the shell, versus its radius normalized
to its r.m.s. radius.  All 16 shells lie virtually on top of one another, 
implying that the shell profile hardly evolves in time. 
However, the profiles that we do not plot do show some evolution.
Those at larger radii evolve as they virialize, 
and those at smaller radii are affected by the finite
resolution of the simulation.
The bottom panel of Figure \ref{fig:shells} shows the slopes
of the profiles in the top panels.

An important question remains: what sets the shell profile shape?
Since the scaled shell profile  is nearly
time-invariant  (Fig. \ref{fig:shells}), and since the 
scale itself is determined by adiabatic shrinking (\S \ref{sec:adshrink}), 
if one could predict the scaled profile just after collapse,
 one could then
predict from first principles the total density profile of
the self-similar solution. 
(In
 \S \ref{sec:toy} we show how to compute the total density profile
 given the shell profile shape.)
We discuss here our preliminary exploration of this question,
  leaving more detailed work to the future. 
 
 A simplistic assumption, useful as a baseline for comparison,
  is to suppose that the
  shell profile is similar to 
 that of a constant density ellipsoid. 
This might be  the case if particle orbits
become highly chaotic after collapse, and if the particles
from each shell uniformly fill an ellipsoidal volume.
A constant density ellipsoid with
semi-axes 
$a\gg b\gg c$ has an 
enclosed mass profile given by 
\be
{d\ln M\over d\ln R}=\cases{
0, & for $R\gg a$ \cr
1, & for $a\gg R\gg b$ \cr
2, & for $b\gg R\gg c$ \cr
3, & for $c\gg R$
}  
\ee
On the smallest lengthscales ($c\gg R$), the enclosed mass
increases as the volume  enclosed by a sphere of radius $R$,
i.e.  $M\propto R^3$.
For comparison, we argued above 
 that in spherically symmetric collapse the shell profile at small $R$ scales as 
$M\propto R$
(\S \ref{sec:s2}), and that in axisymmetric collapse
 $M\propto R^2$
(\S \ref{sec:axi}).  By reasoning similar to that presented in 
\S \ref{sec:axi}, one would conclude that in the triaxial case
the shell profile should scale as $M\propto R^3$ for $R\rightarrow 0$ if particles
uniformly explore a region in three dimensions.   
For a constant density ellipsoid, on scales larger
than $c$ the enclosed mass increases less and less rapidly with $R$ 
as $R$ crosses the semi-major axes of the ellipsoid, until at the largest
 radii
the ellipsoid contributes no more mass, and hence $M\propto R^0$.


In Figure \ref{fig:shells} as well, one can see that the logarithmic
slope of $M$ rolls over from 0 at large radii, to (nearly) 3
at small radii.
The vertical dashed lines show the ratio of the r.m.s. co-ordinates
to
$R_s$, multiplied by $\sqrt{5}$;  that
factor is to account for the fact that  a constant density ellipsoid
has r.m.s. co-ordinates equal to $1/\sqrt{5}$ times its semi-axes.
Hence the profile is broadly similar to that of a constant densitiy 
ellipsoid, with 
the features
in 
$d \ln M/d\ln R$
corresponding to the r.m.s. co-ordinates.  Nonetheless, there
are a number of differences, as discussed below.

\begin{figure}
\centerline{\includegraphics[width=0.5\textwidth]{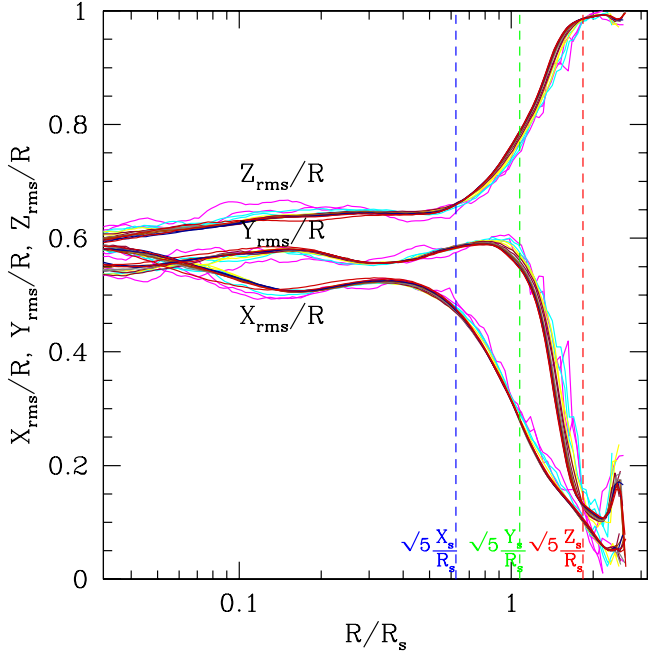}}
\caption{\label{fig:shellaxes}
Triaxiality of Shell Profiles ($\gamma=0.25$, $e=0.1$):
For each of the shell profiles from Figure \ref{fig:shells}, we plot
the r.m.s. co-ordinates of particles that have the same
$R$.  At large radii, most of the particles lie near the $z$-axis, 
while at small radii, the r.m.s. co-ordinates are nearly isotropic. 
This behavior is similar to that of 
a constant density triaxial ellipsoid with semi-axes
$\sqrt{5}X_s$, $\sqrt{5}Y_s$, and $\sqrt{5}Z_s$.
}
\end{figure}
Figure \ref{fig:shellaxes} shows directly the triaxiality
of the shell profiles.  For each shell profile, we plot
the r.m.s. co-ordinates of particles with the same $R$.
At large $R/R_s$ most particles lie along the $z$-axis.
Proceeding to smaller $R/R_s$, particles first approach
isotropy in the $y$ and $z$ directions, and then in all three
directions. At small radii, the three r.m.s. co-ordinates approach
$R/\sqrt{3}=0.577R$, corresponding to isotropy in all three directions.
This behavior is similar to that of a constant density triaxial ellipsoid.

While  Figures \ref{fig:shells}-\ref{fig:shellaxes} indicate that
the shell profile is roughly consistent with 
that of a constant density ellipsoid, 
the agreement is not perfect.
For example, the slope
 $d \ln M/d \ln R$ 
does not quite reach 3 until very small $R$.  
The reason for this appears to be that particles
that initially lie near the $z$-axis (i.e., the eventual
long axis of the ellipsoid) collapse more in the transverse
($x-y$ dimension) than those that lie further from the $z$-axis.
Therefore, the collapsed ellipsoid is in fact more centrally
concentrated than a constant density ellipsoid. However, 
quantifying this effect remains a topic for future work.  
Nonetheless, we recall that the deep shell profile
(at $R\ll R_s$) has little effect on the final density profile
(Fig. \ref{fig:notails}).  Hence it is possible that the details
of this extra concentration are of little importance for the total 
density profile.  

It is instructive to compare our 
 shell profiles with 
those predicted from spherical self-similar models
that include non-radial motions
\citep[e.g.,][]{Nusser01,luetal06,delpopolo09,ZB10}.
In the simplest of these models, particles
are assigned a constant angular momentum (in scaled units)
 at turnaround.   Therefore the shell
 profile is similar to the spherical solution outside
 of periapse, i.e. $M(R)\propto R$, 
for $R_{\rm peri}<R<R_{\rm apo}$, and $M=0$ for 
other values of $R$.
This profile differs from that of Figure \ref{fig:shells}
where the slope of $d\ln M/d\ln R$ gradually rolls over.
More sophisticated spherical infall models have been proposed
that, for example, assign a distribution of angular momenta 
at turnaround \citep{luetal06} or that vary the angular momentum
after turnaround based on a prescription  motivated by tidal
torque theory \citep{ZB10}. 
One could in principal assign the angular momenta
(or equivalently the periapses) in such models
in such a way to mimic our Figure \ref{fig:shells}, or indeed
any shell profile shape.
 Nonetheless, it does not appear
 that such models adequately explain our results, for at least two reasons.
First, when we examine the angular momenta
of individual particles in our solutions, we  find that they 
vary substantially---even over the course of a single orbit---because
of the asphericity of the potential.
By contrast, in spherical models the angular 
momenta are constant  (aside from externally imposed changes). 
In fact, the majority of orbits in our solutions
appear to be more box-like than loop-like. 
This leads us to suppose that our model based on a homogeneous
ellipsoid is more appropriate than spherical models that
are based on angular momentum conservation.
And second, it is evident from Figure \ref{fig:shellaxes}
 that 
triaxiality is playing a leading role in the range of radii
where $d\ln M/d\ln R$ rolls over, whereas in the spherical
models, all three r.m.s. axes would be equal to $0.577R$.
Nonetheless, our explorations at this
stage are preliminary; more self-similar solutions should
be examined, and in more detail,
before definitive statements can
be made regarding what sets the shell profiles.

\section{Toy Model}

\label{sec:toy}

We have shown that two effects are largely responsible for
 the total density profile:  adiabatic shrinking and the
shape of the shell profiles.  Here we construct a simple
toy model that incorporates both effects, similar to 
the models of  
\cite{luetal06, delpopolo09} .
We
 assume that after a shell collapses it has a (scaled) outer
radius $R_{\rm shell}(s)$, and that 
its shell profile is time-invariant when scaled to $R_{\rm shell}$.
This is roughly true of the shell profiles seen in the full simulations.
We define $M_{\rm shells}(R)$ to be the total (scaled) mass
from all shells with $R_{\rm shell}<R$.
Then the total enclosed (scaled) mass is
\be
M(R) = M_{\rm shells}(R) + \int_R^1 {dM_{\rm shells}\over dR'}\mu(R/R')dR' \ ,
\label{eq:toy1}
\ee
where $\mu(R/R')$ is the shell mass profile, i.e. the fraction of mass that
a shell with outer radius $R'$ contributes to a sphere of radius $R$, where
$R<R'$.
Our assumption that the shell profile is invariant when scaled to its outer
radius dictates that $\mu$ is a function only of the ratio $R/R'$; note
that $\mu(R/R')\leq 1$, with equality holding at $R/R'=1$.
Our second assumption is that a shell's proper outer radius 
shrinks adiabatically in inverse proportion to the enclosed mass, i.e.
\be
RM=M_{\rm shells}^{(4+\gamma)/3} \label{eq:toy2}
\ee
(eqs. [\ref{eq:mi}] and [\ref{eq:mrs}]).
Equations (\ref{eq:toy1})-(\ref{eq:toy2}) are the toy model equations.
We drop order-unity constants in these equations, 
which corresponds to setting $R=1$ at the outer-most radius, 
i.e.,  the radius at which
$M=M_{\rm shells}$.
The equations may be solved upon specification
of the shell profile function $\mu()$.
In fact, when 
\be
\mu(R/R')=\left(R/R'\right)^\eta \ , \label{eq:powerlaw}
\ee
 they may
be solved analytically, after re-writing equation (\ref{eq:toy1})
as $dM/dR=(\eta/R)(M-M_{\rm shells})$, and then inserting
equation (\ref{eq:toy2}).

\begin{figure}
\centerline{\includegraphics[width=0.5\textwidth]{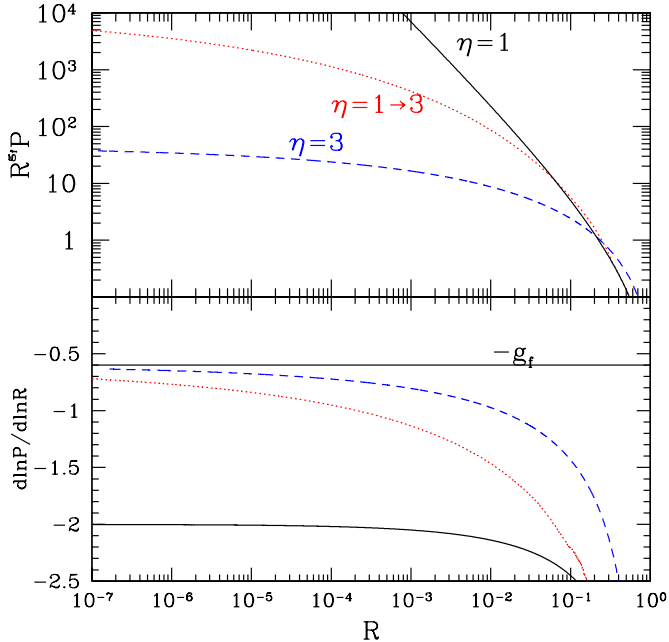}}
\caption{\label{fig:toys}
Toy Model ($\gamma=0.25$):
The top panel shows the compensated density profiles 
that result from solving the toy model equations
(eqs. [\ref{eq:toy1}]-[\ref{eq:toy2}]).
The lower panel shows the logarithmic slopes.
The black solid line is for $\eta=1$ in equation (\ref{eq:powerlaw}), 
corresponding to the shell profiles of the
spherically symmetric solution that have density $\propto R^{-2}$. 
The resulting density profile is also $\propto R^{-2}$.  The
blue dashed line is for $\eta=3$, corresponding to a constant
density sphere.  This is the minimal tail expected, and
leads to a total density profile that rolls over to the frozen
slope over an extended range in $R$.  The red
dashed line is for a shell profile that transitions from $\eta=1$
at large $R/R'$
to $\eta=3$ below $R/R'=1/3$. The resulting total profile
is similar to that seen in the full simulation (Fig. \ref{fig:noshrink}).
}
\end{figure}
Figure \ref{fig:toys} shows the solution for three
possible shell profiles, when $\gamma=0.25$.
The upper panel shows the total density profiles
(compensated by the frozen slope), and the lower panel
shows their logarithmic derivatives.
The blue dashed curve is the result when $\mu=(R/R')^3$. 
 This form  for $\mu$ would result from a constant
density sphere.  It is the minimal ``tail'' expected, since
realistic tails tend to be more centrally concentrated.  
One might have expected that the density in this
 minimal tail model would reach the frozen slope fairly
 quickly.  But in fact it takes over 5 decades in $R$
 to roll over; 
 nonetheless, it does roll over faster than the actual
  solutions
(e.g., Fig. \ref{fig:noshrink}).
The black solid line in Figure \ref{fig:toys} shows 
the result when $\mu=(R/R')$.  This corresponds
to the spherical case, where the shell density profiles
are proportional to $1/R^2$ (\S \ref{sec:sss}).
Since in this case the shell density profiles are steeper
than the frozen slope $g_f=0.6$, the total density profile
asymptotes to that of an individual shell, i.e. to a logarithmic
slope of -2. 
The red dotted curve shows the result of setting
$\mu$ to be a broken power-law, with $\mu\propto(R/R')$
at large radii $(1/3<R/R'<1)$, and $\mu\propto (R/R')^3$ 
at small radii $(R/R'<1/3)$.  This form was chosen as a rough model
for the shell profiles in Figure \ref{fig:noshrink} (see also
Fig. \ref{fig:shells}).
The resulting density profile is quite similar in form to that seen in the 
full simulation of Figure \ref{fig:noshrink}.

In the latter broken power-law model, the roll-over in the shell profile extends
for only half a decade in $R$.  Yet the resulting roll-over in the total 
density profile extends over five decades in $R$.
Recalling that the outer shell profile controls
the total density profile in the self-similar solutions 
(Fig. \ref{fig:notails}), and that the outer shell profile
is in turn largely controlled by the 
ellipticity of collapsed
shells (Figs. \ref{fig:shells} and \ref{fig:shellaxes}), we therefore
attribute the extended roll-over of the total density 
profiles largely 
to the ellipticity of collapsed shells. 
Our picture therefore differs from that of spherical 
infall models with non-radial velocities \cite[e.g.,][]{ZB10}, where it is
 the distribution
of periapses that controls the shell profile, and hence the
total density profile (see also \S \ref{sec:shellprof}).
Our picture suggests an 
explanation 
for why increasing $e$ (the ellipticity of the linear density field)
extends the range in $R$ over which the slope rolls over
towards the frozen slope, 
as shown in Figures \ref{fig:rhos}-\ref{fig:rhosgam1}.
Increasing $e$ makes collapsed shells
more elliptical.  This extends the rollover in the shell profile, 
which in turn extends the rollover of the total density profile.

\section{Summary and Discussion}

We constructed self-similar solutions that describe
the formation, virialization, and growth of dark matter
halos in three dimensions.  
As long as the initial linear density perturbation 
in a flat CDM Universe
can be 
written as $\delta_{\rm lin}\propto r^{-\gamma}f(\theta,\phi)$, 
the subsequent evolution is described by the
self-similar equations
(eqs. [\ref{eq:rv}], [\ref{eq:p}], and [\ref{eq:poiss}]),
 subject to no approximation.
In \S \ref{sec:ssc}, we detailed the algorithms we developed to
solve these equations numerically.
Our solutions are the natural extension of the spherically
symmetric solutions 
 of \cite{FG84} and \cite{Bert85}, and
 of the axisymmetric solutions of
  \cite{Ryden93}, to the case without any assumed
angular symmetry.
Whereas other authors have considered 3D
solutions by modelling the effects of angular momentum, 
our approach
differs, in that we solve the full equations of motion.  This 
leads to more complicated solutions, but
ones that are physically realistic.

Even though our assumed form for $\delta_{\rm lin}$ 
is idealized, the subsequent  self-similar evolution 
can be studied in great detail.
Significantly
higher radial resolution can be achieved than 
with N-body simulations. 
For example, Figure \ref{fig:rhoplot} extends to below
$\sim 10^{-5}$ of the virial radius, whereas the highest
resolution N-body
simulations   reach $\sim 0.004$ of the virial 
radius \citep[e.g.,][]{NLSetal10}.
Furthermore, for any assumed $\gamma$
and angular function $f(\theta,\phi)$, there
is a well-defined {\it solution}, which can be approached
with simulations of ever increasing
 accuracy.\footnote{It is possible that a given  $\delta_{\rm lin}$ will
admit multiple self-similar solutions, i.e., various
nonlinear solutions could be consistent with the same 
linear field.   Although we have not seen evidence 
of this---for example, when we make small changes
in $\delta_{\rm lin}$, the nonlinear solution also changes
by a small amount---it remains a possibility to be explored in
the future.
} 
By contrast, cosmological N-body simulations are initialized
with random fields, and this element of randomness
complicates the analysis of the physical processes
involved. 

In this paper, we used the similarity solutions
 to explore the
processes responsible for the density profile.
In \S \ref{sec:frozen}, we introduced the frozen model, 
in which particles freeze when they reach the
radius of nonlinearity, $r_*$.  This is not meant as a physical model, but
merely to help in the analysis of the full nonlinear solutions.
In \S \ref{sec:sss}, we reviewed the well-known spherical solutions, 
focusing on the processes important for the density profile:
adiabatic contraction and shell profiles, i.e. the density profile
laid down by the set of particles that occupy the same thin
spherical shell at early times.
In \S \ref{sec:axi}, we showed that axisymmetric solutions always
have interior logarithmic slope  $\leq -1$.  
But even though this is
suggestive of NFW,  real halos
are not axisymmetric but triaxial.
In \S \ref{sec:results} we presented a suite of triaxial solutions, 
and analyzed them in considerable detail.
We found that 
the shape of the density profile can be simply understood
by decomposing it into its component shell profiles.
 Our principal results 
are as follows:
\begin{itemize}
\item The density profile rolls over to the frozen slope $-g_f$.  But
the roll-over can extend for many decades in radius.
\item The reason for the extended roll-over can be attributed primarily
to two effects:  adiabatic shrinking and shell tails
\item adiabatic shrinking: the r.m.s. radius of a collapsed shell decreases
in inverse proportion to the enclosed mass (Fig. \ref{fig:shrink}).  A small amount of adiabatic
shrinking can have a surprisingly large influence on the
 density profile (Fig. \ref{fig:noshrink})
\item shell tails:  although the density profile of a collapsed shell extends for 
many orders of magnitude in radius, only 
the outermost decade or so affects the total 
density profile (Fig. \ref{fig:notails}). 
The triaxial shape of a collapsed shell 
is largely responsible for the shell profile 
in this range of radii
(Fig. \ref{fig:shells}, bottom panel and Fig. \ref{fig:shellaxes}).
Furthermore, the 
shape of the shell profile shows little temporal evolution
after virialization (Fig. \ref{fig:shells}, top panel).  
\end{itemize}

In \S \ref{sec:toy}, we incorporated the above results into a toy model,
and showed that
when the shell profile was chosen to approximate that in the self-similar
solution, 
the toy model roughly reproduced the self-similar solution's
total density profile.

The one element missing from our toy model is an estimate
of the shell profile shape given the properties of the linear overdensity.
This is a topic for future work.

In a companion paper \citep{Paper2}, we apply our understanding of what
sets the density profiles in the self-similar solutions to realistic 
halos.
 For such halos, $\delta_{\rm lin}$ is a Gaussian random field, in
which peaks are not scale-invariant and contain a hierarchy of peaks
 within peaks, leading to copious halo substructure.  Despite these
 complications, we show that many aspects of halo structure in
 realistic hierarchical cosmologies may be understood using the
 relatively simple picture described here.

We anticipate that 3D self-similar solutions will be a helpful
tool for exploring the physics of halo formation.
Here we focused on the density profile.  
But the similarity solutions can also be used to explore other
properties of halos, such as the velocity structure within halos, 
the orbital characteristics of collapsed particles, the phase
space density, and the detailed structure of caustics.
These remain topics for future work.

\section*{acknowledgements}
We thank L. Widrow
for useful discussions, and for bringing \cite{Ryden93} to our attention.

\appendix

\section{}
\centerline{APPENDIX \ref{sec:ssc}: NUMERICAL SOLUTION OF THE
SELF-SIMILAR EQUATIONS}
\label{sec:ssc}

In \S \ref{sec:mos} we briefly describe how we solve the
 self-similar equations (eqs. [\ref{eq:rv}], [\ref{eq:p}], and [\ref{eq:poiss}]).
Here we provide more detail.  As a check of our numerical implementation, 
we wrote two independent codes, one by each author;
 one  code employs spherical harmonics
for the angular dependences of the fields, and the other employs 
a  $\theta-\phi$ grid.
We focus first on the former code, and then describe
the principle differences of the latter.

In the spherical harmonic code, the 
density is expanded as
\be
P({\bld R})=\sum_{l,m} P_{lm}(R)Y_{lm}(\theta,\phi) \ , 
\ee
and similarly for $\Phi({\bld R})$.
We assume eightfold
symmetry (i.e., $P(X,Y,Z)=P(-X,Y,Z)$ and similarly for $Y$ and $Z$), and hence
include only terms with even $l$ and  $m$, and $0\leq m\leq l_{\rm max}$.
The functions $P_{lm}(R)$ and $\Phi_{lm}(R)$ are stored on a logarithmic
grid in $R$, with
 around 100 gridpoints per decade in $R$, and
a total extent of 10-20 decades.  

For the first step of each iteration, $P$ is transformed into $\Phi$
by inverting Poisson's equation
(eq. [\ref{eq:poiss}]):
\be
\Phi_{lm}(R)=-{4\pi\over 2l+1}\left(
{1\over R^{l+1}}\int_0^R P_{lm}(R')R'^{l+2}dR'
+R^l\int_R^\infty P_{lm}(R')R'^{-l+1}dR'
\right) \ .
\label{eq:inverse}
\ee
Since the second integral extends to infinity, the linear 
density field (eq. [\ref{eq:linP}]) is used for the part of the integral that
lies beyond the $R$-grid.  For $l=0$, this integral can diverge, in which case
  an infinite constant may be added (which does not affect the equation
of motion) by replacing the second term in parentheses
with -$\int_0^RP_{lm}(R')R'dR'$.

For the second step, particle trajectories are integrated
in a fixed potential (eq. [\ref{eq:rv}]), and their mass is accumulated into
$P_{lm}(R)$.  Particles are initialized at large $s$ where
they nearly expand with the Hubble flow 
$({\bld R}\simeq {\bld s})$.  Their directions
 are chosen 
to be uniformly spaced 
in $\cos\theta_s$ and $\phi_s$,  where these
angles refer to the direction of $\bld{s}$, which differs from the
 direction of $\bld{R}$
by a finite (but small) amount.  Their initial $R$ is chosen to coincide with the last
element of the $R$-grid.  These constraints are satisfied to linear order by
applying the linear solution (\S \ref{sec:lin}), in which
 the linear potential 
comes from the inverse Laplacian of the
imposed linear density (eq. [\ref{eq:linP}]).
The linear solution is also used
to set the initial velocity field.
We employ around 400 particles in $\theta_s$ and 100 in $\phi_s$.

Particle orbits are integrated with 4th-order Runge-Kutta
with adaptive stepping.   The force ${\bld\nabla_R}\Phi$ 
 is evaluated
by spherical harmonic expansion.  The required
$\Phi_{lm}$'s are obtained by linearly interpolating from the grid and
the $Y_{lm}$'s and their derivatives are evaluated at the particle's angular position.
 Also needed
are the $d\Phi_{lm}/dR$; for this purpose we compute a grid of $d\Phi_{lm}/dR$
at the start of each iteration step by evaluating the derivative of the right-hand side
of  equation (\ref{eq:inverse}) on the $R$-grid.

At each $s$ step, the particle's
mass is deposited into $P_{lm}(R)$ according to  equation (\ref{eq:p}).
To be specific, when taking a step $ds$, the right-hand side of equation (\ref{eq:p})
implies that the particle should deposit the (scaled) mass 
$dM=s^2ds d\Omega_s/6\pi$
onto the grid, where $d\Omega_s=d\cos\theta_sd\phi_s=4\pi$/(no. of particles)
 is the 
solid angle subtended in $\bld{s}$.
Therefore this $dM$ is multiplied by the $Y_{lm}^*$'s at the angular position 
of the particle, and added into $P_{lm}$ (with the appropriate weighting
for the size of the relevant $R$-grid element).
To increase the accuracy of the mass deposition, we limit the stepsize $ds$ so that
a particle does not change its $R$ by much more than the local grid spacing.
In addition, whenever a particle crosses a gridpoint in $R$, we use linear
interpolation to find where in $s$ it crossed the gridpoint, and split its deposited
mass accordingly into the two grid bins. This is especially important during
the linear phase of the evolution (at large $R$) where small errors in the
total density produce large errors in the overdensity.  

A particle's integration is terminated when the value of $R$ at its last apoapse
is comparable to the $R$ of the first grid element.  Note that at
 late times apoapses
typically remain constant or shrink in $r$ (i.e. in real, not scaled coordinates).
Therefore, their $R$ decreases with decreasing $s$ in proportion to $s^{\gamma+1}$, 
or faster.  Hence all particles eventually have apoapses with arbitrarily small $R$.
In addition to this stopping criterion, we also occasionally limit $s$ to a minimum
value, or limit the number of apoapses per particle.

The algorithm as described above is trivially parallelized.
We have done this by allocating
a subset of the particles to each processor.
 Each processor
holds a copy of the $\Phi_{lm}$, and accumulates the $P_{lm}$ from its own
particles.  At the end of each iteration step, the $P_{lm}$ from the different
processors are then combined and inverted to yield the $\Phi_{lm}$.

 We have also written a second independent code to solve the
self-similar equations, as a check of our numerical solutions.
 Both codes use similar particle integration schemes and radial
 resolution, but differ principally in the implementation of the
 force solver.  Instead of computing spherical harmonic
 coefficients $P_{lm}$ directly from the particle data, our second
 code bins the density into a three-dimensional grid, in $\log R$,
 $\theta$ and $\phi$.  Then, the spherical harmonic coefficients
 are computed using the fast transform library 
 {\tt S2kit}.\footnote{\tt http://www.cs.dartmouth.edu/$\sim$geelong/sphere}
 We then solve for the potential $\Phi$ and its derivatives on the
 grid, using spherical harmonic transforms, and interpolate the
 force from the 3-D grid onto the particle positions as required
 for the orbit integrations.
 
 Our discrete sampling of the initial phase space can lead to spurious
 high-$l$ power in the linear regime at large $R$.  To suppress this
 effective shot noise, we use the high-order Triangle-Shaped Cloud
 interpolation \citep{HockneyEastwood} for the angular grid, and simple
 linear interpolation in the radial direction.  In addition, because
 the angular grid does not have a pixel at $\theta=0$, special care
 must be taken to handle particles passing near the pole.  We keep an
 additional array storing the potential derivatives at $\theta=0$ for
 each radial bin, calculated analytically from the harmonic
 coefficients, and employ it in the force interpolation for particles
 traversing the smallest $\theta$ grid cells. 
  The memory requirements of this code are minimal, and 
 the inter-node communications are negligible, leading to efficient
 performance on essentially any architecture.  For a typical run with
 $l_{\rm max}=28$, radial range of 16 decades in $R$, and 
 ${\cal O}(10^7)$ particles, each iteration requires roughly one hour
 on 60 nodes, and the solutions reach convergence in typically 5-10
 iterations, though we usually ran 15 iterations to be assured of
 convergence.

\section{}
\centerline{APPENDIX \ref{sec:lin}: LINEAR SOLUTION OF THE SELF-SIMILAR
EQUATIONS}
\label{sec:lin}

The self-similar equations ([\ref{eq:rv}], [\ref{eq:p}], and [\ref{eq:poiss}]) 
have zeroth order solution (i.e., in the limit $R\rightarrow \infty$):
\be
{\bld R_0} = {\bld s} ; \ {\bld V_0} = {2\over 3}{\bld s} ;\  \Phi_0 = {1\over 9}R^2 ; \ P_0 = {1\over 6 \pi} \ .
\ee
To linear order, we set
\be
P_1(\bld{R}) = {1\over 6\pi}R^{-\gamma}f(\theta,\phi) \ ,
\ee
where $f$ is arbitrary.
The linear potential is
\be
\Phi_1(\bld{R}) = R^{2-\gamma}g(\theta,\phi)  \ ,
\ee
where $g$ may be found from $f$ by inverting Poisson's equation (eq. [\ref{eq:poiss}]).
Equation (\ref{eq:rv}) is to linear order
\be
{d\over d\ln s}
\left(\begin{array}{c}{\bld R_1} \\ {\bld V_1}\end{array}\right)=
\left(\begin{array}{cc}
\gamma+1
& -{3\gamma\over 2} \\
{\gamma\over 3} & 1-{\gamma\over 2}\end{array}\right)
\left(\begin{array}{c}{\bld R_1} \\ {\bld V_1}\end{array}\right) 
+
{3\gamma\over 2}
\left(\begin{array}{c}0 \\ {\bld \nabla_s\Phi_1}\end{array}\right) \ ,
\ee
considering $\Phi_1$ to be a function of $\bld{s}$ instead of
$\bld{R}$, which is valid to linear order.
Because this is a linear equation, $\bld{R_1}, \bld{V_1}$ must
be constants times 
 $\bld{\nabla_s}\Phi_1$, and hence they 
 scale with $s$ as $s^{1-\gamma}$. Replacing
 ${d\over d\ln s}\rightarrow 1-\gamma$  gives
\beqn
\bld{R_1}& =& -{3\over 2}\bld{\nabla_s}\Phi_1 \\
\bld{V_1}&=&-2\bld{\nabla_s}\Phi_1 \ .
\eeqn
 The velocity relative to the local Hubble flow is
$\bld{V}-2\bld{R}/3=-{\bld\nabla_s}\Phi_1$, a well-known result.
  The linearized equation describing mass deposition (eq. [\ref{eq:p}]),
\be
6\pi P_1 ={d^3s\over d^3R}-1 
=-{\bld\nabla_s\cdot R_1(s)} = {3\over 2}\nabla^2_s\Phi \ , \label{eq:mdlin}
\ee
is equivalent to Poisson's equation.  The reason for this redundancy is that
in writing equation (\ref{eq:rv}) we already took into account the linear evolution
in setting the form of $r_*$.  One could alternatively set $r_*=t^{\alpha}$,
with arbitrary $\alpha$; equation (\ref{eq:mdlin}) would then prove that 
$\alpha={2\over 3}+{2\over 3\gamma}$.

\section{}
\centerline{APPENDIX \ref{sec:ep}: PARAMETERIZATION OF THE LINEAR
DENSITY FIELD}
\label{sec:ep}

We follow \citet{BondMyers96} in parameterizing the local tides near peaks using the ellipticity $e$ and prolateness $p$ of the tidal field, defined as follows.  Let $\Phi$ be the peculiar gravitational potential smoothed on scale $R_{\rm smooth}$, and write $\{\lambda_a\},\ a=1,2,3$ as the eigenvalues of the tidal tensor $\nabla_i\nabla_j\Phi$, ordered such that $\lambda_1 \leq \lambda_2 \leq \lambda_3$.  Note that the overdensity $\delta=\sum_a \lambda_a$.   The ellipticity and prolateness are then defined as
\begin{eqnarray}
e &=& \frac{\lambda_3-\lambda_1}{2\delta} \\
p &=& \frac{\lambda_3-2\lambda_2+\lambda_1}{2\delta}.
\end{eqnarray}

In general, $e$ and $p$ are explicit functions of the smoothing scale.  For our scale-free initial profiles, however, the smoothing scale cancels, and $e$ and $p$ may be expressed in terms of the relative amplitudes of the $l=2$ multipoles compared to the monopole.  If we write the linear density profile as
\begin{equation}
\rho(r,\theta,\phi) = r^{-\gamma} [1 + a_{20} Y_{2,0}(\theta,\phi) + a_{22} (Y_{2,2} + 
Y_{2-2})]
\end{equation}
then we have
\begin{eqnarray}
a_{20} &=& \sqrt{10\pi} (3e+p) \frac{\gamma}{3-\gamma} \label{eq:a20}\\
a_{22} &=& \sqrt{\frac{15\pi}{2}} (e-p) \frac{\gamma}{3-\gamma}
\label{eq:a22}
\end{eqnarray}

\bibliographystyle{hapj}
\bibliography{ms}

\begin{thebibliography}{21}
\expandafter\ifx\csname natexlab\endcsname\relax\def\natexlab#1{#1}\fi

\bibitem[{{Ascasibar} {et~al.}(2004){Ascasibar}, {Yepes}, 
  {Gottl{\"o}ber},\& M{\"u}ller   }]{Ascasibar04}
{Ascasibar}, Y., {Yepes}, G.,  {Gottl{\"o}ber}, S., \& {M{\"u}ller}, V. 2004, \mnras, 352, 1109,
  arXiv:astro-ph/0312221

\bibitem[{{Ascasibar} {et~al.}(2007){Ascasibar}, {Hoffman}, \&
  {Gottl{\"o}ber}}]{Ascasibar07}
{Ascasibar}, Y., {Hoffman}, Y., \& {Gottl{\"o}ber}, S. 2007, \mnras, 376, 393,
  arXiv:astro-ph/0609713

\bibitem[{{Bardeen} {et~al.}(1986){Bardeen}, {Bond}, {Kaiser}, \&
  {Szalay}}]{BBKS}
{Bardeen}, J.~M., {Bond}, J.~R., {Kaiser}, N., \& {Szalay}, A.~S. 1986, \apj,
  304, 15

\bibitem[{{Bertschinger}(1985)}]{Bert85}
{Bertschinger}, E. 1985, \apjs, 58, 39

\bibitem[{{Bond} \& {Myers}(1996)}]{BondMyers96}
{Bond}, J.~R., \& {Myers}, S.~T. 1996, \apjs, 103, 1

\bibitem[{{Dalal} {et~al.}(2010){Dalal}, {Lithwick}, \& {Kuhlen}}]{Paper2}
{Dalal}, N., {Lithwick}, Y., \& {Kuhlen}, M. 2010, ArXiv e-prints, 1010.2539

\bibitem[{{Del Popolo}(2009)}]{delpopolo09}
{Del Popolo}, A. 2009, \apj, 698, 2093, 0906.4447

\bibitem[{{Diemand} {et~al.}(2007){Diemand}, {Kuhlen}, \& {Madau}}]{ViaLactea1}
{Diemand}, J., {Kuhlen}, M., \& {Madau}, P. 2007, \apj, 667, 859,
  arXiv:astro-ph/0703337

\bibitem[{{Diemand} {et~al.}(2008){Diemand}, {Kuhlen}, {Madau}, {Zemp},
  {Moore}, {Potter}, \& {Stadel}}]{ViaLactea2}
{Diemand}, J., {Kuhlen}, M., {Madau}, P., {Zemp}, M., {Moore}, B., {Potter},
  D., \& {Stadel}, J. 2008, \nat, 454, 735, 0805.1244


\bibitem[{{Duffy} \& {Sikivie}(1984)}]{DuffySikivie}
{Duffy}, L.~D., \& {Sikivie}, P. 2008, \prd, 78, 063508


\bibitem[{{Eisenstein} \& {Loeb}(1995)}]{el}
{Eisenstein}, D.~J., \& {Loeb}, A. 1995, \apj, 439, 520


\bibitem[{{Fillmore} \& {Goldreich}(1984)}]{FG84}
{Fillmore}, J.~A., \& {Goldreich}, P. 1984, \apj, 281, 1

\bibitem[{{Gao} {et~al.}(2008){Gao}, {Navarro}, {Cole}, {Frenk}, {White},
  {Springel}, {Jenkins}, \& {Neto}}]{Gao08}
{Gao}, L., {Navarro}, J.~F., {Cole}, S., {Frenk}, C.~S., {White}, S.~D.~M.,
  {Springel}, V., {Jenkins}, A., \& {Neto}, A.~F. 2008, \mnras, 387, 536,
  0711.0746

\bibitem[{{Hockney} \& {Eastwood}(1988)}]{HockneyEastwood}
{Hockney}, R.~W., \& {Eastwood}, J.~W. 1988, {Computer simulation using
  particles} (Bristol: Hilger)

\bibitem[{{Lu} {et~al.}(2006){Lu}, {Mo}, {Katz}, \& {Weinberg}}]{luetal06}
{Lu}, Y., {Mo}, H.~J., {Katz}, N., \& {Weinberg}, M.~D. 2006, \mnras, 368,
  1931, arXiv:astro-ph/0508624

\bibitem[{{Moore} {et~al.}(1998){Moore}, {Governato}, {Quinn}, {Stadel}, \&
  {Lake}}]{Moore98}
{Moore}, B., {Governato}, F., {Quinn}, T., {Stadel}, J., \& {Lake}, G. 1998,
  \apjl, 499, L5+, arXiv:astro-ph/9709051

\bibitem[{{Navarro} {et~al.}(2010){Navarro}, {Ludlow}, {Springel}, {Wang},
  {Vogelsberger}, {White}, {Jenkins}, {Frenk}, \& {Helmi}}]{NLSetal10}
{Navarro}, J.~F. {et~al.} 2010, \mnras, 402, 21, 0810.1522

\bibitem[{{Nusser}(2001)}]{Nusser01}
{Nusser}, A. 2001, \mnras, 325, 1397, arXiv:astro-ph/0008217

\bibitem[{{Ryden}(1993)}]{Ryden93}
{Ryden}, B.~S. 1993, \apj, 418, 4

\bibitem[{{Ryden} \& {Gunn}(1987)}]{RydenGunn87}
{Ryden}, B.~S., \& {Gunn}, J.~E. 1987, \apj, 318, 15

\bibitem[{{Stadel} {et~al.}(2009){Stadel}, {Potter}, {Moore}, {Diemand},
  {Madau}, {Zemp}, {Kuhlen}, \& {Quilis}}]{GHalo}
{Stadel}, J., {Potter}, D., {Moore}, B., {Diemand}, J., {Madau}, P., {Zemp},
  M., {Kuhlen}, M., \& {Quilis}, V. 2009, \mnras, 398, L21, 0808.2981

\bibitem[{{Vogelsberger} {et~al.}(2010){Vogelsberger}, {Mohayaee}, \&
  {White}}]{VMW10}
{Vogelsberger}, M., {Mohayaee}, R., \& {White}, S.~D.~M. 2010, ArXiv e-prints,
  1007.4195

\bibitem[{{White} \& {Zaritsky}(1992)}]{WhiteZaritsky92}
{White}, S.~D.~M., \& {Zaritsky}, D. 1992, \apj, 394, 1

\bibitem[{{Zukin} \& {Bertschinger}(2010)}]{ZB10}
{Zukin}, P., \& {Bertschinger}, E. 2010, ArXiv e-prints, 1008.0639

\end{thebibliography}

\end{document}